\documentclass[10pt,pre,aps,onecolumn,superscriptaddress,nobibnotes,nodoi,noeprint]{revtex4-2}

\usepackage{silence}
\usepackage{graphicx}
\usepackage{dblfloatfix}
\usepackage{amsmath,physics,bm,float}
\usepackage{soul}
\usepackage[colorlinks=true,citecolor=blue,urlcolor=blue,linkcolor=blue]{hyperref}
\usepackage{amssymb}
\usepackage{natbib}
\usepackage{upgreek}
\usepackage{mathtools}
\usepackage{appendix}
\usepackage{color,xcolor}
\usepackage{multirow}
\usepackage{bm}

\usepackage[mathlines]{lineno}
\let\oldequation\equation\let\oldendequation\endequation
\renewenvironment{equation}{\linenomathNonumbers\oldequation}{\oldendequation\endlinenomath}
\let\oldalign\align\let\oldendalign\endalign
\renewenvironment{align}{\linenomathNonumbers\oldalign}{\oldendalign\endlinenomath}

\newcommand{\hhu}{Institut f\"ur Theoretische Physik II: Weiche Materie, Heinrich-Heine-Universit\"at D\"usseldorf, Universit\"atsstra{\ss}e 1, D-40225 D\"usseldorf,
	Germany}
 \newcommand{\tud}{Institut f\"ur Physik der kondensierten Materie, Technische Universit\"at Darmstadt, Hochschulstra{\ss}e 8, D-64289 Darmstadt, Germany}
\newcommand{\jgu}{Institut f\"ur Physik, Johannes Gutenberg-Universit\"at Mainz, 
55128 Mainz, 
Germany}
\newcommand{\sau}{Theoretische Physik, Universit\"at des Saarlandes, Campus E26, D-66123 Saarbr\"ucken, Germany
}

\begin{document}

\title{
Anomalous Mean-Squared Displacement in Quantum Active Matter from a Wigner Phase-Space Framework
}

\author{Sangyun Lee}
\email{sanlee@uni-mainz.de}
\affiliation{\jgu}

\author{Yehor Tuchkov}
\affiliation{\jgu}

\author{Alexander P.\ Antonov}
\affiliation{\hhu}

\author{Benno Liebchen}
\email{benno.liebchen@pkm.tu-darmstadt.de}
\affiliation{\tud}

\author{Hartmut L\"owen}
\email{Hartmut.Loewen@uni-duesseldorf.de}
\affiliation{\hhu}

\author{Giovanna Morigi}
\affiliation{\sau}
\affiliation{Center for Quantum Technologies (QuTe), Saarland University, Campus, 66123 Saarbr\"ucken, Germany}

\author{Michael te Vrugt}
\email{tevrugtm@uni-mainz.de}
\affiliation{\jgu}

\date{\today}
\begin{abstract}
Active matter is driven out of equilibrium by a local influx of energy. While classical active matter has been extensively studied, the extension of active matter concepts to quantum systems has been explored far less. In this work we develop a quantum description based on the Wigner function for a quantum system influenced by a classical noise. 
For the model considered here, the hybrid dynamics reduces to an Ornstein--Uhlenbeck process, enabling us to compute the quantum mean-squared displacement (MSD) using established techniques from classical active matter.
We analytically derive the time dependence of the MSD and clarify the conditions under which the characteristic scaling with time $\mathrm{MSD}\sim t^{6}$ emerges, namely the regime of long persistence time and large active noise strength. 
We also show that, for certain parameter and initial conditions, the MSD can exhibit an even steeper scaling regime $\mathrm{MSD}\sim t^{7}$.
In addition, explicit expressions are derived that precisely predict the onset times of $t^6$ and $t^7$ scaling behaviors.
Finally, we examine the robustness of these behaviors against quantum fluctuations of the initial state.
\end{abstract}

\maketitle

\section{Introduction}

Active matter \cite{ramaswamy2010mechanics, marchetti2013hydrodynamics, gompper20202020, Schuettler2025Active} refers to nonequilibrium systems consisting of units that continuously convert internal or environmental energy into directed motion. 
Across micro- and macroscales, these systems display a wide range of nonequilibrium phenomena, 
from ballistic motion at the single-particle level \cite{Szamel2014Self,Bechinger2016Active} to spontaneous self-organization at the collective level \cite{vicsek1995novel, toner1995long, soto2014self, schmidt2019light,kalz2024field}. 
Typical examples of active matter include living organisms \cite{cavagna2014bird, ramaswamy2017active} and robots \cite{baconnier2022selective, caprini2024emergent, antonov2024inertial} on the macroscale and bacteria \cite{aranson2022bacterial} or Janus particles \cite{walther2008janus, liebchen2018synthetic} on the microscale. 
Although the smallest active objects, such as enzymes \cite{jee2018catalytic, ghosh2021enzymes}, can be as small as several nanometers, this is still far beyond the
scale where the quantum effects typically emerge. 
However, recent advances in the control of individual atoms in the quantum regime have made the experimental realization of active particle exhibiting quantum behavior conceivable, 
raising the fundamental question for how to define quantum active matter~\cite{vrugt2025exactly}.

Quantum active matter refers to active matter whose behavior is affected by quantum effects such as interference, entanglement, quantum statistics and discretization of energy levels. 
It is an emerging research area and, so far, only a few studies have addressed quantum active matter, including hopping-based lattice models~\cite{adachi2022activity,takasan2024activity,khasseh2025active,penner2025heat}, nonreciprocal spins~\cite{nadolny2025nonreciprocal} and models for diffusion in continuous space~\cite{antonov2025engineering,antonov2025modeling} An experimental implementation has also recently been explored~\cite{BurgardtEtAl2026}. One approach~\cite{takasan2024activity} is to model quantum active matter using spin systems. In particular, a lattice model of hard-core bosons with spin-dependent asymmetric hopping, which acts as an activity, has been shown to exhibit activity-induced ferromagnetism.
Waveguide systems have been proposed as a platform for implementing quantum active matter and realizing nonreciprocal phase transitions~\cite{nadolny2025nonreciprocal}.
Another approach~\cite{antonov2025engineering} combines a quantum system with a classical active system, allowing the quantum system to inherit activity from the classical dynamics. 
Within this framework, the mean-squared displacement (MSD) -- a widely studied quantity in classical active matter physics \cite{grossmann2024non,lemaitre2023non} -- was investigated, An unusual 
$t^6$ scaling of the MSD of a quantum active particle was reported in Ref.~\cite{antonov2025engineering}. 
This scaling is anomalous in the context of active matter, where the MSD of systems such as run-and-tumble particles and the active Ornstein–Uhlenbeck process typically exhibits a ballistic $t^2$ regime at short times and crosses over to diffusive scaling, $\sim t$, at long times.

In quantum systems, a single point in phase space does not correspond to a realizable event in the classical probabilistic sense, due to the noncommutativity of quantum observables. Specifically, the Heisenberg uncertainty principle forbids a simultaneous sharply localized initial distribution in both position and momentum. 
Nevertheless, phase-space representations such as Wigner transformation~\cite{breuer2002theory,gardiner2004quantum} remain useful, particularly for extending classical systems to the quantum regime, where intuition from classical dynamics plays a central role~\cite{dekker1977quantization}. 
Furthermore, in some systems, this representation even allows the use of classical analytical or computational tools to analyze quantum dynamics.
Therefore, in the context of quantum active matter, which combines intrinsically nonequilibrium driving with quantum fluctuations, such a representation can be valuable. 

In this paper, we further investigate the condition and robustness of the anomalous $t^6$ scaling reported for the MSD of a system mimicking quantum active matter~\cite{antonov2025engineering}. 
The numerical analysis revealing the $t^6$ scaling in Ref.~\cite{antonov2025engineering} was based on the long-time-limit expression for the quantum MSD and a quantum master equation valid in small activity regime. 
Here we extend this analysis by evaluating the MSD without the long-time approximation and by considering an open quantum master equation valid beyond small activity regime~\cite{antonov2025modeling}. In addition, we choose the initial condition such that the activity
contributes already at $t=0$, allowing us to capture the influence of activity from the earliest stage of the dynamics.

To this end, we employ a hybrid Wigner master equation to analytically evaluate the MSD of the system mimicking quantum active matter. 
Compared to the previous studies~\cite{antonov2025engineering,antonov2025modeling}, our framework enables the derivation of an exact analytical expression for the MSD, allowing us to systematically characterize its scaling behavior and identify the regimes in which the $t^6$ scaling emerges.
We systematically characterize the behavior of the MSD and find the regime where the $t^6$ scaling emerges.
In addition to finding the emergence condition of the $t^6$ scaling, we find that for the initial condition that is considered in earlier work~\cite{antonov2025modeling} the MSD can also exhibit a $t^7$ scaling regime in the limit of strong active diffusion. 
Finally, we examine the robustness of these anomalous scalings with respect to the choice of the initial quantum state. 

\section{ Wigner transformation } 

Let us first outline the general idea of this paper: To investigate the behavior of the MSD, we introduce a hybrid Wigner master equation that handles classical and quantum randomness on a phase space. As the hybrid Wigner master equation has the same mathematical structure as the classical Fokker--Planck equation, the hybrid master equation allows us to employ established tools from classical active matter, for example to calculate the MSD of the underlying quantum system. This provides an efficient way to analyze the quantum active matter. Note that we use the term {\it hybrid} for denoting that a joint quasiprobability of state variables for quantum and classical systems, and we do not consider backaction effect of a quantum state on a classical state, which is included in dynamics discussed in Ref.~\cite{Diosi2023Hybrid}.

The Wigner transformation~\cite{Wigner1932}  provides a useful tool that  allows a quantum system to be represented in phase space in a manner analogous to that of a classical system. This representation enables a comparison between classical and quantum dynamics~\cite{Lee2020Finite,Lee2021Quantumness}, and also offers insights into quantum extensions of classical theories, as early works on open quantum systems~\cite{dekker1977quantization}.
The Wigner function is not a probability distribution, since individual points in phase space do not correspond to realizable physical events. 

The Wigner transform of an operator $\hat A$ is given by
\begin{align}
A_W(x,p)
= {\Xi}[\hat A]
= \frac{2}{\hbar}
\int dz \,
e^{- \frac{2 i p z}{\hbar}}
\left\langle x+z \middle| \hat A \middle| x-z \right\rangle 
\label{eq:AW}.\end{align}
Here $\hbar$ is the reduced Planck's constant and $\hat \cdot $ denotes quantum operator.
Here, $x$ and $p$ represent the phase-space variables of the quantum system. For classical systems, a subscript $c$ is used to distinguish their state variables from those of the quantum phase space.
For the density matrix $\hat{\rho}$, the corresponding Wigner function reads 
\begin{align}
W(x,p,t)
\equiv& \frac{1}{2\pi}{ \Xi} [ \hat\rho(t)]
\end{align} where the factor $1/(2\pi)$ is required for normalization.

The dynamics of an open quantum system, such as the Lindblad master equation~\cite{breuer2002theory,Rivas2012open}, can also be described in terms of the Wigner function. Assuming a quantum master equation with system Hamiltonian~$\hat H$ and a dissipator $\mathcal D_{\rm diss}$, the master equation is written as 
\begin{align}
    \frac{ d \hat{ \rho }(t)  }{ d t} = -\frac{i}{\hbar}[ \hat H ,\hat\rho (t)  ]+{\mathcal D}_{\rm diss}[\hat \rho(t) ], 
\label{eq:basic_QME}
\end{align}
In our work, we consider a Lindblad dissipator~\cite{breuer2002theory} for $\mathcal D_{\rm diss}$.
After the Wigner transformation~\cite{groenewold1946principles,breuer2002theory,gardiner2004quantum}, Eq.~\eqref{eq:basic_QME} is given by 
\begin{align}
    \partial_t W(x,p) =& H_W \star W - W \star H_W + D_{diss} [W]\label{eq:wignermas_pre}\\
    \equiv& \mathcal L_W W(x,p)
    \label{eq:wignermas}
\end{align}
where $H_W = {\Xi}[\hat H]$ is the Wigner transformed Hamiltonian and $\star$ denotes the Moyal product (See Ref.~\cite{groenewold1946principles} for more information on Moyal product.) 
\begin{align}
        (f \star g)(x,p)
    \equiv
    f(x,p)\,
    \exp\!\left[
    \frac{i\hbar}{2}
    \left(
    \overleftarrow{\partial_x}\overrightarrow{\partial_p}
    -
    \overleftarrow{\partial_p}\overrightarrow{\partial_x}
    \right)
    \right]
    g(x,p)
.\end{align}
Here, $\overleftarrow{\partial_y}$ ($\overrightarrow{\partial_y}$) denotes the derivative operator acting on the function to its left (right), respectively.
In Eq.~\eqref{eq:wignermas_pre}, $ H_W \star W - W \star H_W $ describes the evolution generated by the Liouville operator~$-(i/\hbar) [\hat H, \hat \rho]$ and $D_{diss} [W]$ denotes dissipation by external environment.
Using computer algebra software such as {\it Mathematica}~\cite{Mathematica}, the transformation can be carried out easily.
We provide our code that can handle the Wigner transformation analytically in Ref.~\cite{Tuchkov2026WolframCode}. 
The operator expectation values can be calculated in terms of the phase-space averaging with $W$. 
For example, the expectation of a symmetrized one-time operator is equivalent to the expectation value of the corresponding one-time function in the Wigner phase space (see appendix~\ref{append:onetimeequiv}). 
Therefore, Eq.~\eqref{eq:wignermas_pre} can be utilized to calculate physical quantities.

We can interpret Eq.~\eqref{eq:wignermas} as a classical Fokker--Planck equation, when the Wigner master equation is quadratic in $\partial_x$ and $\partial_p$ and the initial condition is a Gaussian function. The mentioned condition ensures that the Wigner function is always positive. 
Therefore, there exists the corresponding Langevin equation to the Wigner master equation, in the same way that a stochastic process governed by a Langevin equation can be equivalently described by a classical Fokker–Planck equation.
This is referred to as a quasiclassical Langevin equation~\cite{gardiner2004quantum}.
Even when the Wigner function becomes negative in certain regions of phase space and the quantum master equation is quadratic to $\hat x$ and $\hat p$, the time evolution of its moments coincides with that obtained from the corresponding classical Fokker–Planck equation and from the associated Langevin equations~\cite{gardiner2004quantum}. 

For example, let us assume that Wigner master equation is quadratic in $\partial_{x}$ and $\partial_{p}$ and that its drift vector and diffusion matrix are given by a linear function $\mathbf f(\mathbf q,t)$ and $BB^T$, respectively. Then, the master equation introduced above corresponds to the quasiclassical Langevin equation 
\begin{align}
    \dot{\mathbf q}(t) = \mathbf f(\bm q,t) + B \cdot \bm{\eta}(t)
\end{align}
where $\langle \eta_i (t) \eta_j (t') \rangle  = 2 \delta_{ij}\delta(t-t')$. 
Note that $\bm q$ is a real-valued vector representing coordinates in Wigner phase space and does not correspond to an operator.
The vector $\mathbf f$ is a linear function of $\mathbf q$. Because nonlinear force terms lead to higher-order derivatives $\partial_{\mathbf{q}}^{\,n}$ $(n>2)$ in the Wigner master equation, they cannot be handled by such a quasiclassical Langevin equation.

In this work, we focus on a heat bath described by a Lindblad master equation~\cite{antonov2025modeling} and determine the MSD scaling with time. Since the $t^6$ scaling~\cite{antonov2025engineering} was identified in the regime where the effect of ${\mathcal D}_{\rm diss}[\hat \rho(t) ]$ is small, the specific choice of quantum heat bath does not affect the conditions for the emergence of the $t^6$ or $t^7$ scalings that we will discuss.
The Lindblad master equation with a harmonic potential centered at $x= x_c$, given by $\hat U(x) = m\omega^2 (\hat x-x_c)^2/2$, where $m$ is the mass of the system and $\omega$ is the frequency of the trap leads to the quasiclassical Langevin equation~\cite{antonov2025modeling}
\begin{subequations}    \label{eq:langevinL}
\begin{align}
    \dot x(t)  &= \frac{p(t)}{m} - \bar\gamma [x(t) - x_c(t)] 
    + B_x \eta_{x}(t),   \label{eq:langevinLpos}\\
    \dot p(t) &= -m\omega^2 [x(t)-x_c(t)]
    - \bar\gamma p(t) + B_p \eta_{p}(t) ,
\end{align}
\end{subequations}
where $\bar\gamma$ and $B_i$ are friction and diffusion coefficients. Note that $x(t)$ and $p(t)$ in Eq.~\eqref{eq:langevinL} are different from $x$ and $p$ in Eq.~\eqref{eq:AW}. The latter are phase-space variables.
The thermostat acting on the positional degrees of freedom contributes to the initial linear behavior of the quantum MSD. We will explain this point in the following section. 
The inhomogeneous damping in Eq.~\eqref{eq:langevinLpos} is unavoidable when one employs a Markovian master equation with a harmonic oscillator satisfying complete positivity and trace preservation (CPTP)~\cite{Massignan2015Quantum}.

Eq.~\eqref{eq:langevinL} belong to the class of Ornstein–-Uhlenbeck processes. This allows us to employ established techniques commonly used in classical active matter physics to analyze the dynamics of the quantum system. In the following, we show how the MSD can be calculated analytically using methods developed for classical systems.

Due to the correspondence of one-time correlation function, the quasiclassical Langevin equation can be utilized to calculate quantum expectation values. 
The MSD is an important quantity that characterizes a system in a wide range of systems from active matter to glassy systems~\cite{sastry1998signatures}. In the Heisenberg picture, it is defined by 
\begin{align}
    \langle [\hat x(t) - \hat x(0)]^2 \rangle_{\hat\rho} = \langle \hat x(t)^2\rangle_{\hat\rho} - \langle \hat x(t) \hat x(0)\rangle_{\hat\rho} - \langle\hat x(0)\hat x(t)\rangle_{\hat\rho} + \langle \hat x(0)^2\rangle_{\hat\rho}   
\label{eq:MSD_main}
\end{align} 
where $\langle \cdot \rangle_{\hat\rho}$ denotes average over a density matrix, $\hat \rho$. 
Although we assume here a one-dimensional system, this definition straightforwardly extends to an $N$-dimensional system. We note that the definition of the MSD used in this work [Eq.~\eqref{eq:MSD_main}] differs from that employed in Refs.~\cite{antonov2025engineering,antonov2025modeling}, where the MSD is defined without involving two-time correlation functions.

The MSD contains a symmetrized two-time correlation function, which can be calculated from the Wigner master equation or the quasi-classical Langevin equation using the following equality
\begin{align}
\langle
\hat x(t)\hat x(0) + \hat x(0)\hat x(t)
\rangle_{\hat \rho}
=& \int dx dp [  x_W(t)\star  x_W(0)  + x_W(0)\star  x_W(t) ]W(x,p,0)
\label{eq:equaility}.\end{align}
When $\hat x(t)$ is a linear function of $\hat x(0)$ and $\hat p(0)$, then the right hand-side of Eq.~\eqref{eq:equaility} reduces to a simpler form, and the two-time correlation functions calculated in terms of the Wigner function are equal to the operator expectation of the symmetrized two-time correlation function.  
\begin{align}
\langle
\hat x(t)\hat x(0) + \hat x(0)\hat x(t)
\rangle_{\hat \rho}
=& 2 \int dx dp \,x_W(t)   x_W(0)  W(x,p,0)
\label{eq:equiv}
\end{align}
Here, taking $\hat A=\hat x,\hat p$ in Eq.~\eqref{eq:AW}, the corresponding Wigner functional is $x_W(t)$ and $p_W(t)$. 
In Appendix~\ref{append:onetimeequiv}, we present a general proof of the equivalence, from which Eq.~\eqref{eq:equiv} follows as a special case. 
Equation~\eqref{eq:equiv} provides a convenient form for calculations, and, in the remainder of this paper, we focus on systems for which this equality holds.

\begin{figure}[b!]
\includegraphics[width=0.48\textwidth]{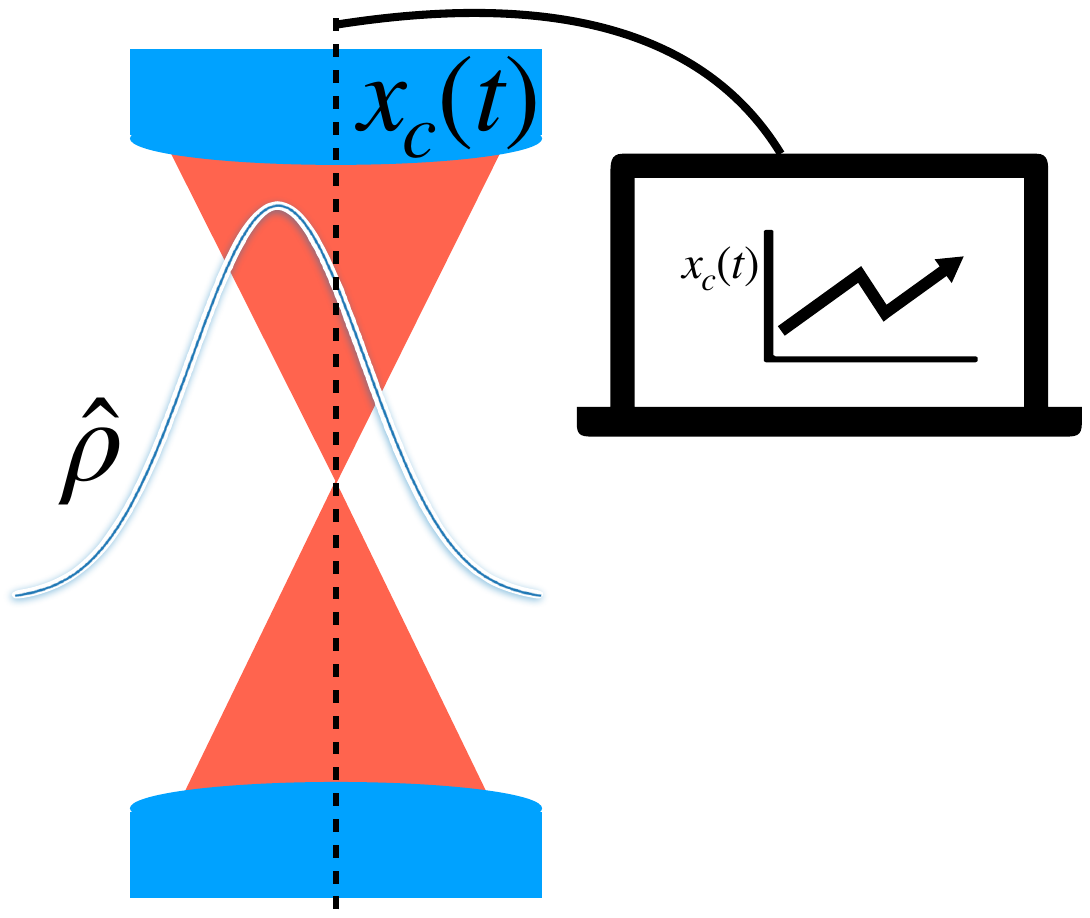}
    \caption{Conceptual figure to illustrate the model mimicking quantum active matter from Ref.~\cite{antonov2025modeling}.
    The Hamiltonian for this system is given by Eq.~\eqref{eq:hamiltonian}. The optical tweezer exerts harmonic force on the quantum system, $\hat\rho$. The computer generates activity term~$x_c (t)$, and the center of the optical trap follows the generated trajectory $\Gamma_c = \{x_c(t)\}$. 
    }
    \label{fig:concept}
\end{figure}

\section{ Hybrid Wigner Master Equation } 

In the context of quantum active matter, one approach~\cite{antonov2025engineering,antonov2025modeling} to extend classical active matter into the quantum regime is to construct systems that combine classical stochastic dynamics with quantum degrees of freedom. 
This scheme is particularly useful for describing systems that mimic quantum active particles. 
Consider an atom trapped by an optical tweezer whose potential can be approximated as a harmonic oscillator, with its center denoted by $x_c(t)$ (see Fig.~\ref{fig:concept} for a schematic illustration). 
The trap center follows a trajectory~$\{x_c(t)\}_{0\leq t}$ generated by numerically solving a classical Langevin equation for active noise. 
Through the movement of the trap center, the quantum state~$\hat \rho$ of the atom inherits the features of classical active matter~\cite{antonov2025engineering, antonov2025modeling}.

The particle that we consider is not an autonomously self-propelling particle. However, the term `active matter' is now very well established both for intrinsically self-propelled particles and for externally actuated particles. In fact, some externally actuated systems, such as granular particles on vibrated plates~\cite{Vijay2007Long}, show the same statistical properties and even follow the same equations of motion as intrinsically self-propelled particles~\cite{Scholz2018Inertial}. So, at a mathematical level, there is no difference between both cases. The same is true for the model explored in Ref.~\cite{antonov2025modeling} and in the present work, which shows, in the classical limit, the same characteristic intermediate- and late-time behavior in the mean-squared displacement as intrinsically self-propelled particles. Still, to emphasize external actuation rather than intrinsic self-propulsion, following Ref.~\cite{antonov2025modeling}, in the present work we use the term ``mimicking" active motion.

Since the potential of the quantum active matter follows the classical stochastic equation, it is natural to introduce a hybrid Wigner function that incorporates classical stochasticity alongside quantum dynamics. 
Let us assume that $\bm q_c$ is the state of a classical system and $\Gamma_c \equiv \{\bm q_c(t)\}$ is its trajectory.
The distribution of $\bm q_c$ evolves according to
\begin{align}
    \partial_t P(\bm q_c, t) = \mathcal L_c P(\bm q_c, t )
.\end{align}
For a given realization $\Gamma_c $, the quantum master equation is given by
\begin{align}
    \frac{d\hat \rho
    (t;\Gamma_c)}{dt} = -\frac{i}{\hbar}[ \hat H (\bm q_c,t) ,\hat\rho (t;\Gamma_c)  ]+{\mathcal D}[\hat \rho(t;\Gamma_c) ], 
\label{eq:hybridQME}\end{align}
and the corresponding Wigner master equation for $\Gamma_c$ is 
\begin{align}
\partial_t W(x,p,t | \Gamma_c)
=
\mathcal{L}_W 
W(x,p,t | \Gamma_c )
\end{align}
Here, $W(x,p,t | \Gamma_c)$ is attained by Wigner transforming the density matrix $\rho_c$ that is conditioned on a given trajectory $\bm q_c$.
For a given trajectory $\Gamma_c$, Eq.~\eqref{eq:hybridQME} provides a CPTP map when the superoperator is written in the Lindbladian form~\cite{breuer2002theory}, which leads to the positivity of the hybrid density matrix conditioned on $\bm q_c$ and is important to the validity of the hybrid Wigner function defined below (see Appendix.~\ref{append:valid} for further details).
Then, we define a hybrid quasi-probability distribution which combines the trajectory probability $P[\Gamma_c]$ and the Wigner function 
\begin{align}
W(x,p,\bm q_c,t)
\equiv
\int \mathcal{D}[\Gamma_c]\,
W(x,p,t | \Gamma_c)\,P[\Gamma_c]\,
\delta \left(\bm q_c - \bm q_c(t)\right)
.\end{align} 
Then, the master equation for the hybrid function is given by 
\begin{align} \label{eq:hybridWignerGeneral}
    \partial_t W(x,p, \bm q_c,t)
=
\mathcal L_W W(x,p,\bm q_c,t)
+
\mathcal L_{FP} W(x,p,\bm q_c,t)
.\end{align}
In Appendix~\ref{append:hybridWigner}, we provide the derivation for the above master equation. Although the derivation in Appendix~\ref{append:hybridWigner} assumes specific stochastic differential equation, the result can be straightforwardly extended to a general stochastic differential equation for $\bm q_c$. Because the mathematical structure of Eq.~\eqref{eq:hybridWignerGeneral} is analogous to that of stochastic equations used in classical active matter, Eq.~\eqref{eq:hybridWignerGeneral}, together with the quasiclassical Langevin equation, may enable the application of tools and analysis familiar to researchers from classical active matter physics. 

As there exists an equivalence between the MSD from quantum operative expectation and the one from Wigner space expectation for a given trajectory, the hybrid function also provides the equivalence even after averaging over realization of the classical trajectory $\bm q_c(t)$. In the following section, we present the application of this equivalence for finding detailed condition for $t^6$ slope of the MSD of quantum active matter that recently discussed in Ref.~\cite{antonov2025engineering} and shows the condition when $t^7$ slope arises. 

\section{Scaling investigation using Hybrid Wigner function}
\label{sec:t7behav}

We consider the following one-dimensional hybrid quantum master equation describing a quantum system coupled to classical variables for activity~$x_c$ and $u$. 
This is the model from Ref.~\cite{antonov2025modeling}, expressed in terms of a quantum master equation. It reads
\begin{align}
    \frac{ d \hat{ \rho }_{ c }(t)  }{ d t} = -\frac{i}{\hbar}[ \hat H(x_c (t) ),\hat\rho_{ c} (t)  ]+{\mathcal D}[\hat \rho_{ c } (t) ], 
\label{eq:hybrid_Lind}\end{align}
where $\Gamma_c=\{x_c(t)\}$ is a given classical trajectory, $\hat \rho_c (t) \equiv \hat \rho (t;\Gamma_c) $
denotes the density matrix of the quantum system conditioned on the trajectory and ${\mathcal D}[\hat \rho_{ c } (t) ]$ is the superoperator describing the effects of a thermal heat bath. Here, the commutator is defined as $[\hat A,\hat B]\equiv \hat A\hat B - \hat B\hat A$. 
Note that when an observable is averaged over random trajectories, the corresponding probability may be expressed as the combination of the density matrix and the trajectory probability $\hat \rho_c(t) P[\{x_c(t)\}]$. For example, in Ref.~\cite{antonov2025engineering,antonov2025modeling}, the average over the combined probability was used to evaluate the approximated MSD.

The first term on the right-hand side of Eq.~\eqref{eq:hybrid_Lind} corresponds to the unitary evolution by the Hamiltonian of kinetic energy and harmonic potential centered at $x_c$:
\begin{align}
	\hat{H}(x_c (t) )=\frac{\hat{p}^2}{2 m} +\frac{1}{2} m \omega^2 \left[\hat{x}-x_c(t)\right]^2
	\label{eq:hamiltonian}
.\end{align}
Here, $x_c(t)$ denotes the internal active variable that encodes the active fluctuations.
The non-Markovian character of $x_c(t)$ gives rise to persistent motion, which, as we will show later, leads to a crossover from ballistic dynamics to enhanced diffusion, a characteristic feature of active-particle dynamics.
Its dynamics is governed by Ornstein--Uhlenbeck process~\cite{antonov2025modeling},
\begin{subequations}
\label{eq:AOUP}
\begin{align}
	\dot{x}_c(t) =& u(t), \label{eq:u_def}\\
    \tau\dot{u}(t) =& -u(t) + \sqrt{D}\eta(t) \\
    =& -u(t) + \tau \sqrt{D_u}\eta(t)
\end{align}
\end{subequations}
where $D$ is a rescaled diffusion coefficient defined in terms of the
diffusion coefficient $D_u$ and the persistence time $\tau$ as $D \equiv \tau^2 D_u $.
$\eta$ is the Gaussian white noise satisfying $\langle \eta(t) \eta(t') \rangle = 2 \delta (t-t') $.
In the long-time limit ($t\gg \tau $), $D$ governs the diffusive behavior of $x_c$, leading to the MSD,
\begin{align}
    \langle x_c^2(t) \rangle_c \simeq 2D\, t 
.\end{align}
In contrast, $D_u$ is the diffusion coefficient for the auxiliary variable~$u$. In the short-time limit, 
\begin{align}
    \langle u^2\rangle_c \simeq 2 D_u t
.\end{align}
$\langle \rangle_c$ denotes the average over the noise realizations in Eqs.~\eqref{eq:AOUP}, representing an averaging over classical-thermal fluctuations. 
The dynamics of the active Ornstein--Uhlenbeck process (AOUP) introduces nonequilibrium fluctuations into the quantum system through $x_c(t)$. $ u(t)$ is an auxiliary variable that can be interpreted as the velocity of $x_c(t)$ and allows the dynamics to be expressed as a Markovian process. From now on, we consider the trajectories of $x_c$ and $u$ rather than $x_c$ alone. 
The second term describes non-unitary evolution and has the Lindblad form
\begin{align}
\label{eq:Lindblad-dynamic}
{\mathcal{D}[\hat\rho]}  \equiv &\frac{\gamma}{2}\bar n\left(  \hat{a}^\dagger (t)\hat{\rho}(t) \hat{a}(t) - \frac{1}{2}\left\{\hat{a}(t)\hat{a}^\dagger(t) ,\hat{\rho} (t)\right\} \right) \nonumber \\
& + \frac{\gamma}{2}(\bar n + 1) \left(  \hat{a} (t)\hat{\rho}(t) \hat{a}^\dagger(t)- \frac{1}{2} \left\{\hat{a}^\dagger (t)\hat{a}(t) ,\hat{\rho}(t) \right\} \right) .
\end{align}
Here, $\bar n = [\exp(\hbar\omega/k_B T)-1]^{-1}$ is the mean number of quanta in equilibrium, and the anti-commutator is defined as $\{\hat A,\hat B\} \equiv \hat A\hat B + \hat B\hat A$. The creation and annihilation operators $\hat{a}^\dagger(t)$ and $\hat{a}(t)$ are defined with respect to the instantaneous position of the potential minimum $x_c(t)$ as 
\begin{subequations}
\begin{align}
\hat{a}(t)  =  \sqrt{\frac{m\omega}{2\hbar}} \left( \hat{x} - {x}_c(t)+ \frac{i}{m\omega} \hat{p} \right)
\end{align}
and
\begin{align}
\hat{a}^\dagger(t)  = \sqrt{\frac{m\omega}{2\hbar}} \left( \hat{x} - {x}_c(t)- \frac{i}{m\omega} \hat{p} \right).
\end{align}
\label{eq:CA1}
\end{subequations}
The parameters $\nu_- = \gamma(\bar n + 1)/2$ and $\nu_+ = \gamma \bar n/2$ denote the cooling and heating rates, respectively. More specifically, $\nu_-$ is the prefactor for the term $\hat{a} (t)\hat{\rho}(t) \hat{a}^\dagger(t)- \frac{1}{2} \left\{\hat{a}^\dagger (t)\hat{a}(t) ,\hat{\rho}(t) \right\}$ in Eq.~\eqref{eq:Lindblad-dynamic}, which reduces the number of quanta in the system. $\nu_+$ is the prefactor for the term $\hat{a}^\dagger (t)\hat{\rho}(t) \hat{a}(t) - \frac{1}{2}\left\{\hat{a}(t)\hat{a}^\dagger(t) ,\hat{\rho} (t)\right\}$ in Eq.~\eqref{eq:Lindblad-dynamic}, which increases the number of quanta in the system. 

Eq.~\eqref{eq:Lindblad-dynamic} is derived under the adiabatic quantum master equation approximation; the change rate of the system Hamiltonian is sufficiently slow so that it does not lead to additional nonadiabatic contributions on $\mathcal D [\hat\rho _c] $~\cite{albash2012quantum, Dann2018Time}. 
For a detailed discussion, see Refs~~\cite{albash2012quantum, Dann2018Time}. 
A criterion for the validity of Eq.~\eqref{eq:Lindblad-dynamic} is given by
\begin{align}
    \ell \sim \sqrt{\frac{mD_u\tau}{2\omega^2\hbar^2} \max{\left(k_BT - \frac{\hbar \omega}{2},0\right)}} \ll 1
\label{eq:criterion_main}.\end{align} In Appendix~\ref{append:criterion}, the derivation of the above criterion is provided.

We now introduce the Wigner representation of the hybrid quantum--classical system.
The Wigner transform of the density matrix of a given trajectory of $x_c(t)$ and $u(t)$ is defined as
\begin{align}
    W_c(x,p,t)
    \equiv 
    \mathcal{W}\!\left[\hat{\rho}_c(t)|
    \Gamma_c\right]
.\end{align} 
Here, the trajectory is written as $\Gamma_c = \{ x_c(t), u(t)\}$.
For a given $\{x_c(t)\}$, the Wigner equation is given by 
\begin{align}
\label{eq:FPE2}
    \partial_t W_c(x,p,t) & = \partial_i[f_i W_c(x,p,t)] + \partial^2_{ij}[g_{ij} W_c(x,p,t)], \nonumber\\
    & \equiv \mathcal L_{W}W_c(x,p,t)], 
\end{align}
where the drift vector and the diffusion matrix are given by 
\begin{subequations}
\begin{align}
    f_i & =  \begin{pmatrix} -\frac{p}{m} + \frac{\gamma}{4}(x-x_c) \\ m\omega(x-x_c) + \frac{\gamma}{4}p\end{pmatrix},
\end{align}
and 
\begin{align}
    g_{ij} & = \begin{pmatrix}
        \frac{\gamma\hbar}{8m\omega}\coth\left(\frac{\hbar \omega}{2k_B T} \right) & 0 \\ 0 & \frac{\gamma\hbar m\omega}{8}\coth\left(\frac{\hbar \omega}{2k_B T} \right)
    \end{pmatrix}
.\end{align}
\end{subequations}
For the derivation of Eq.~\eqref{eq:FPE2} see Ref.~\cite{antonov2025modeling}.
Here, we introduce the hybrid Wigner function such that 
\begin{align}
    W(x,p,x_c,u,t)
    =
    \int \mathcal D [\Gamma_c] W_c(x,p,t)P[\Gamma_c] \delta (x_c -x_c(t) )\delta(u - u(t)).
\end{align}
Then, by following a derivation analogous to that of the differential Chapman–Kolmogorov equation~\cite{van1992stochastic,gardiner2009stochastic}, the master equation for the hybrid Wigner function is written by  
\begin{align}
    \partial_t W(x,p,x_c,u,t) = \mathcal L_w W(x,p,x_c,u,t) + \mathcal L_{\rm FP} W(x,p,x_c,u,t)
\end{align}
where $\mathcal L_{\rm FP}$ is the generator for the evolution of the AOUP. In Appendix~\ref{append:hybridWigner}, we present the detailed derivation. 
The Fokker--Planck equation for the AOUP [Eqs.~\eqref{eq:AOUP}] reads
\begin{align}
    \partial_t P(x_c, u,t) =& (\partial_{x_c}, \partial_u ) \cdot 
    \left(\begin{smallmatrix}
        0 & -1 \\
        0 & \frac{1}{\tau}
    \end{smallmatrix}
    \right) \cdot \left(\begin{smallmatrix}
    x_c \\
    u
    \end{smallmatrix}\right)  P(x_c,u,t)\nonumber\\
    &+ (\partial_{x_c}, \partial_u ) \cdot 
    \left(\begin{smallmatrix}
        0 & 0 \\
        0 & D_u
    \end{smallmatrix}
    \right) \cdot \left(\begin{smallmatrix}
    \partial_{x_c}\\
    \partial_u
    \end{smallmatrix}\right)  P(x_c,u,t) \\
    \equiv& \mathcal L_{FP} P(x_c,u,t)
,\end{align} 
which is straightforwardly derived from the classical Langevin equation, Eq.~\ref{eq:AOUP}.

Therefore, the master equation for the hybrid Wigner function is given by this hybrid Fokker--Planck equation
\begin{align}
    \partial_{t} W(x,p,x_c,u,t) =& 
    (\partial_{x},\, \partial_{p},\, \partial_{x_c},\, \partial_{u} ) 
    \cdot
    \left(\begin{smallmatrix}
    \frac{\gamma}{4} & -\frac{1}{m} & -\frac{\gamma}{4} & 0 \\
    m\omega^2 & \frac{\gamma}{4} & -m\omega^2  & 0 \\
     0 & 0 & 0 & -1 \\
     0& 0 & 0 & \frac{1}{\tau} 
    \end{smallmatrix}\right)
    \cdot
    \left(\begin{smallmatrix}
    x\\
    p \\
    x_c \\
    u
    \end{smallmatrix}\right)W(x,p,x_c,u,t)
    \nonumber\\
    &+ 
    \left(\partial_{x},\, \partial_{p},\, \partial_{x_c},\, \partial_{u} \right) 
    \cdot
    \left(\begin{smallmatrix}
    \frac{\gamma\hbar}{8m\omega}\coth\left(\frac{\hbar \omega}{2k_B T} \right) & 0& 0& 0  \\
    0 & \frac{\gamma\hbar m\omega}{8}\coth\left(\frac{\hbar \omega}{2k_B T} \right)& 0& 0\\
    0 & 0& 0& 0\\
    0 & 0& 0& D_u\\
    \end{smallmatrix}\right)
    \cdot
    \left (\begin{smallmatrix}
    \partial_{x}\\
    \partial_{p}\\
    \partial_{x_c}\\
    \partial_{u}
    \end{smallmatrix}\right)W(x,p,x_c,u,t)\\
    =& \mathcal L_{\rm hyb} W(x,p,x_c,u,t)
.\end{align}
   
Since the hybrid Wigner master equation has a Fokker–Planck structure, there can be many corresponding stochastic differential equations~\cite{gardiner2009stochastic}. These stochastic differential equations describe the evolution of a classical probability distribution governed by $\partial_t P(x,p,x_c,u,t) = \mathcal{L}_{\rm{hyb}} P(x,p,x_c,u,t)$.
Because both the hybrid Wigner dynamics and the associated stochastic process are generated by the same operator $\mathcal{L}_{\mathrm{hyb}}$, the time evolution of all moments are equivalent in the two descriptions. 

We can write the corresponding quasiclassical Langevin equation as
\begin{align}
    \dot{\mathbf q}_t = - A \cdot \mathbf q_t + B \cdot \bm{\eta}(t),
\label{eq:OULangevin}\end{align}
where 
\begin{align}
A =
\left(\begin{smallmatrix}
\frac{\nu_- - \nu_+}{2} & -\frac{1}{m} & -\frac{\nu_- - \nu_+}{2} & 0 \\
m\omega^2 & \frac{\nu_- - \nu_+}{2} & -m\omega^2 & 0 \\
0 & 0 & 0 & -1 \\
0 & 0 & 0 & \frac{1}{\tau}
\end{smallmatrix}\right)
,\,
B =
\left(\begin{smallmatrix}
\sqrt{\frac{\hbar(\nu_+ + \nu_-)}{4m\omega}} & 0 & 0 & 0 \\
0 & \sqrt{\frac{m\omega\hbar(\nu_+ + \nu_-)}{4}} & 0 & 0 \\
0 & 0 & 0 & 0 \\
0 & 0 & 0 & \sqrt{D_u}
\end{smallmatrix}
\right)
,
    \mathbf q(t) =
    \left (\begin{smallmatrix}
    {x(t)}\\
    {p(t)}\\
    {x_c(t)}\\
    {u(t)}
    \end{smallmatrix}\right)
,\end{align}
\,\text{ and }
\begin{align}
    \bm{\eta}(t) =
    \left (\begin{smallmatrix}
    {\eta_x(t)}\\
    {\eta_p(t)}\\
    {0}\\
    {\eta_u(t)}
    \end{smallmatrix}\right)
.\end{align}
Here, $\eta_i$ is a Gaussian white noise with zero mean and $\langle \eta_i (t)\eta_j (t') \rangle = 2\delta_{ij}\delta(t-t')$. With the Langevin equation, the analytical form for the MSD can be derived (see Appendix~\ref{append:solveCorr}). 
The stochastic differential equation consists of the linear drift term and Gaussian white noise term, which is named Ornstein-Uhlenbeck process~\cite{gardiner2009stochastic}. 
The covariance matrix of $\mathbf  q(t)-\mathbf q(0)$ is given by
\begin{align}
\langle (\mathbf q(t) -\mathbf q(0) ) \otimes (\mathbf q(t) -\mathbf q(0) ) \rangle =&
e^{-A t} \langle \mathbf q (0) \otimes \mathbf q(0)\rangle  e^{-A^T t}  
+  2\int_0^t e^{-A(t - s)} B B^T e^{-A^T(t - s)} ds  \nonumber\\
&-  e^{-A t} \langle \mathbf{q}(0)\otimes \mathbf{q}(0)\rangle -  \langle \mathbf{q}(0)\otimes \mathbf{q}(0)\rangle  e^{-A^T t} 
+\langle \mathbf q (0) \otimes \mathbf q(0) \rangle 
.\label{eq:Cov}
\end{align}
Then, the MSD is given by one of the diagonal components of Eq.~\eqref{eq:Cov} as
\begin{align}
    \langle |\mathbf x(t) - \mathbf x(0) |^2\rangle =\left[\langle (\mathbf q(t) -\mathbf q(0) ) \otimes (\mathbf q(t) -\mathbf q(0) ) \rangle\right]_{xx}
    \label{eq:MSD}
.\end{align}
Due the aforementioned equivalence, evaluating Eq.~\eqref{eq:MSD} provides the MSD of the hybrid system,
\begin{align}
    \int \mathcal D[\Gamma_c] \langle [\hat x (t) -\hat x(0)]^2 \rangle_{ \Gamma_c}
    P[\Gamma_c]= \left[\langle (\mathbf q(t) -\mathbf q(0) ) \otimes (\mathbf q(t) -\mathbf q(0) ) \rangle\right]_{xx}
,\end{align}
where $\langle \rangle_{ \Gamma_c }$ denotes the quantum average conditioned on a given trajectory $\{x_c(t)\}$.
Note that when numerically evaluating the MSD [Eq.~\eqref{eq:MSD}], we solve an ordinary differential equation rather than directly evaluating the integral in Eq.~\eqref{eq:Cov}. Both approaches are numerically indistinguishable (see Appendix~\ref{append:evaluateTwoTimeFunction}). 
\subsection{MSD with the initial condition $P(x_c, u)=\delta(x_c)\exp{[- u^2/(2D_u \tau) ]}/\sqrt{2\pi D_u\tau}$}

\begin{figure}[htbp]
\includegraphics[width=0.96\textwidth]{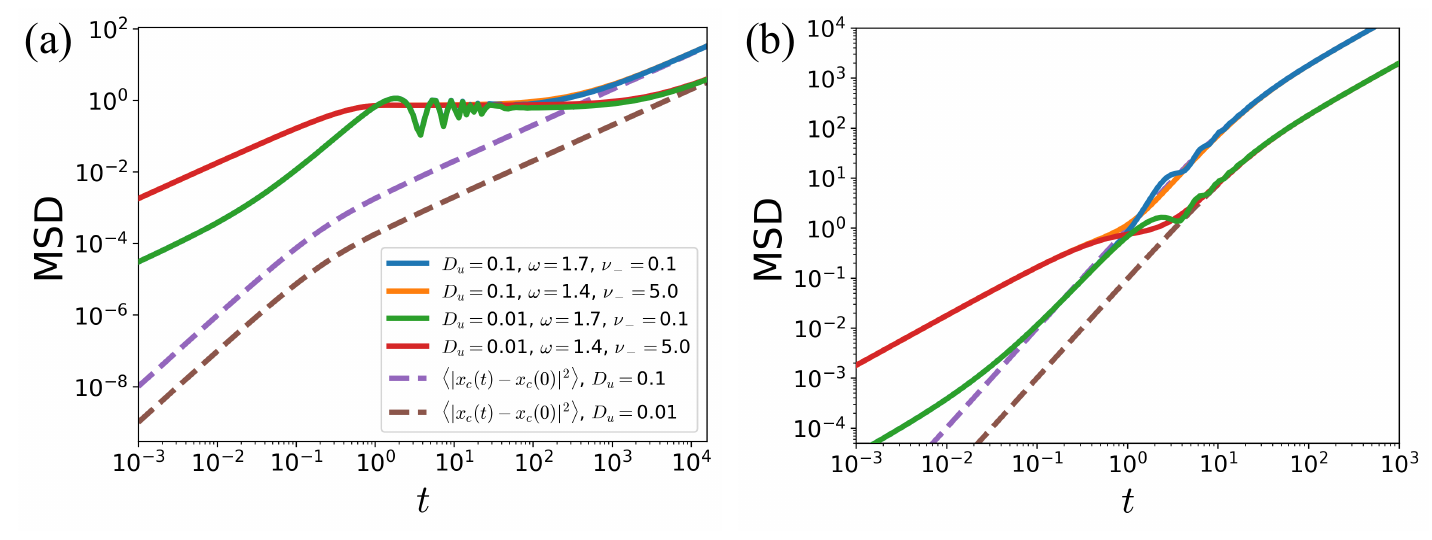}
\vskip -0.1in
\caption{Plot of MSD versus time of the quantum active particle with the initial conditions,~Eq.~\eqref{eq:inicond1}. We varied $D_u$, $\omega$, $\nu_-$, and $\nu_+$ while fixing the ratio $\nu_-$ to $\nu_+$ as $10^4$. Two persistence times were considered: (a) $\tau =0.1$ case (b) $\tau =10$ case. The solid lines represent the MSDs with each parameter combinations. The dashed lines represent the MSD of $x_c(t) $. The same color coding is applied to both figures. All quantities in this and the following figures are shown in arbitrary units. 
}
    \label{fig:main}
\end{figure}

We consider the initial condition of joint distribution of ground state $|0\rangle$ of the harmonic oscillator centered at $x=0$, while 
\begin{align}
    P(x_c,u) = \delta(x_c) \exp{[- u^2/2D_u\tau]}/\sqrt{2\pi D_u\tau}
.\label{eq:inicond1}\end{align} 
With this initial condition, the dynamics reduces to that of a classical AOUP with colored noise from the initial time and its MSD is given by
\begin{align}
    \langle |x_c (t) - x_c (0)|^2 \rangle_c =2 D_u \tau^2 [t + \tau(-1 + e^{ - t/\tau  }) ]
\label{eq:usualAOUPMSD}.\end{align} Under this initial condition the MSD of $x_c$ shows scaling transition from $t^2$ to $t$. The derivation is same for the Brownian diffusion case (See Section 2 of Ref.~\cite{Qirezi2009}).
Note that this condition is different from the one in Refs~\cite{antonov2025engineering, antonov2025modeling}. The latter case will be considered in the following subsection. The system has three characteristic time scales: the persistence time $\tau$, the inverse of the harmonic frequency $1/\omega$, and the inverse of the dissipation rate $1/\gamma$. This can be seen in Eq.~\eqref{eq:Cov}, which implies that the eigenvalues of $A$ govern the time dependence. The eigenvalues of $A$ are given by $\lambda_1 =0$ $\lambda_2 = 1/\tau$, $\lambda_3 = \gamma/4 - i\omega  $ and $\lambda_4 = \gamma/4 + i\omega $. 
These eigenvalues define the characteristic rates of the dynamics.
$\lambda_3$ and $\lambda_4$ indicate that the MSD oscillates with a high value of $\omega$ and a small value of $\gamma$.

\begin{figure}
\includegraphics[width=0.48\textwidth]{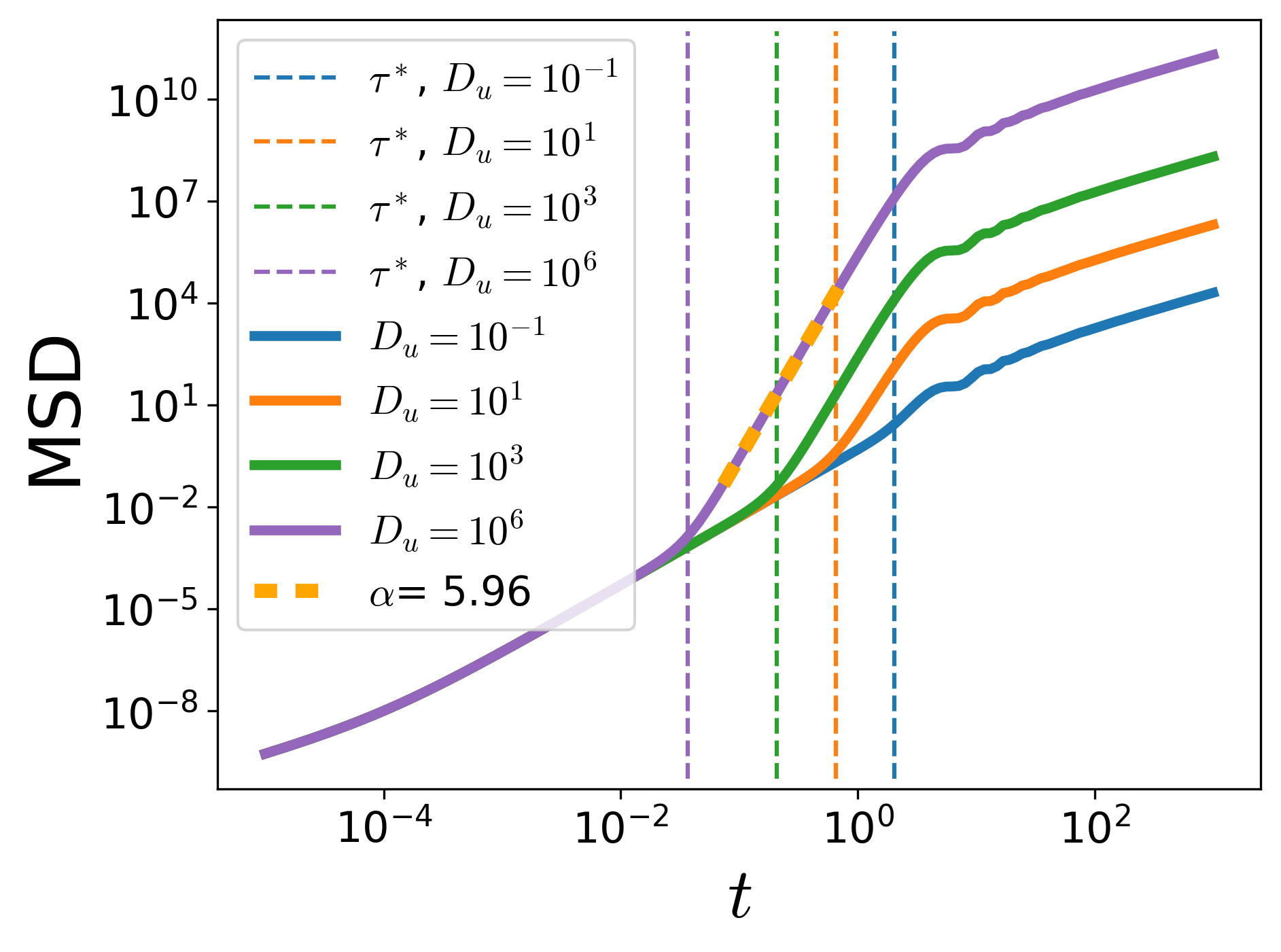}
\vskip -0.1in
\caption{ Plot of MSD versus time of the quantum active particle with different $D_u$. $D_u$ is varying from $0.1$ to $10^6$. Vertical lines represent the threshold time, Eq.~\eqref{eq:taustar}.
Other parameters are chosen as follows: $\omega = 1.0$, $\hbar = 1.0$, 
$\nu_- = 10^{-4}$, $\nu_+ = 10^{-8}$, and $\tau = 10$. 
The dashed line represents $t^\alpha$ where $\alpha = 5.9577...\pm 0.00296$. 
 We set the mass and Planck constant to unity.
}
\label{fig:t6activity}
\vskip -0.1in
\end{figure}
%)

\begin{table}[h]
\centering
\begin{tabular}{|c|c|c|c|c|c|}
\hline
&  very early & early & intermediate & late  & $t \gg \tau$ \\
\hline
Long $\tau$ with moderate $\nu_-$ and $\nu_+$& \multicolumn{2}{|c|}{$t$} & $t^\alpha$, $\alpha < 6$ or oscillatory & { $t^2$ }&{$t$} \\
\cline{1-4}
Long $\tau$, weak $\nu_-$ and $\nu_+$, high $D_u$ &  $t$ & $t^2$ &$t^\alpha$, $\alpha \simeq 6$ or oscillatory &  &  \\
\hline
Short $\tau$ ($<1/\omega$, $1/\gamma$) with moderate $\nu_-$ and $\nu_+$ & \multicolumn{2}{|c|}{$t$} &  $t^\alpha$, $\alpha < 6$ & \multicolumn{2}{|c|}{$t$} \\
\cline{1-3}
Short $\tau$ with weak $\nu_-$ and $\nu_+$&  $t$ & $t^2$ & or oscillatory  & \multicolumn{2}{|c|}{} \\
\hline
\end{tabular}
\caption{MSD of a quantum active particle for different times and parameter regimes. 
We choose the initial condition for classical degrees of freedom as $P(x_c, u)=\delta(x_c)\exp{[- u^2/(2D_u \tau) ]}/\sqrt{2\pi D_u\tau}$.
With this choice, the scaling of $\langle|\hat x-x_c|^2 \rangle $ is bounded from above by $\mathrm{MSD}\sim t^6$ in the intermediate-time regime. The system exhibits a crossover from persistent motion including ballistic motion (MSD $ \sim t^2$) to enhanced diffusion ($\sim t$). 
}
\label{table1}
\end{table}

In Fig.~\ref{fig:main}, we present two cases (a) $\tau=0.1$ and (b) $\tau=10$. One can see that the system transitions from persistent motion, including ballistic motion, to enhanced diffusion, which is a typical feature of classical and quantum active particles~\cite{Gipouloux2026Active}.
In Fig.~\ref{fig:main} (a), $\tau$ is the shortest time scale among the three characteristic time scales. We consider the two possible orderings 
$1/\omega > 1/\gamma > \tau$ and $1/\gamma > 1/\omega > \tau$. 
For each ordering, we examine two values of $D_u$: a large value ($D_u = 1000$) and a small value ($D_u = 0.1$). In Fig.~\ref{fig:main} (b), $\tau$  is the longest, or at least comparable to the other characteristic time scales. 
As in Fig.~\ref{fig:main} (a), we consider the same two orderings of time scales, 
$1/\omega > 1/\gamma > \tau$ and $1/\gamma > 1/\omega > \tau$, again for both $D_u = 1000$ and $D_u = 0.1$.

In both cases, the initial dynamics is primarily governed by $\omega$, 
whereas the late behavior is determined by the value of $D_u$. This can be understood from the fact that $\omega$ determines the initial variance of position and momentum. The late behavior can be understood the fact that the MSD of the quantum active matter approaches to that of $x_c$,
\begin{align}
    \langle |\hat x(t) - \hat x(0)|^2 \rangle \sim \langle |x_c (t) -x_c(0)|^2 \rangle_c \text{ when $t \rightarrow \infty$}
,\end{align} and the MSD of $x_c$ transitions from $t^2$ to $t$. 
When $\tau$ is the shortest time scale, the MSD converges to that of $x_c$ after the latter enters $t$ scaling. Conversely, when $\tau$ is the longest time scale, the MSD of our quantum active matter converges to that of $x_c$ once the latter exhibits ballistic $t^2$ scaling.

Under the assumption that $\tau$ is the longest timescale, we derived the asymptotic form of the MSD
\begin{align}
    \langle |\hat x(t) - \hat x(0)|^2 \rangle \sim 2D_u\tau^2 t + B_0
\Big[ 
-g_2(t) e^{-2t/\tau}
+g_1(t) e^{-t/\tau} 
- g_0
\Big],
\label{eq:aysmptotic}\end{align} where $g_1(t)$ and $g_2(t)$ are linear function of $t$ (see Appendix~\ref{appen:asymptotic} for its derivation and explicit forms of coefficients). Therefore, two terms involving exponential decaying factor will be suppressed after $t=\tau$. 
Eq.~\eqref{eq:aysmptotic} implies that after $\tau$ the MSD approaches to linear regime as $2D_u\tau^2 t$.

In the intermediate regime, various dynamical features appear. If $\gamma=2(\nu_--\nu_+)$ is small, the MSD oscillates. When $\gamma$ is high, the oscillatory behavior vanishes and the scaling of MSD is $t^\alpha$ where $\alpha\leq 6$. Depending on $D_u$ and $\omega$, a decrease of the MSD is also possible, which can be found from Fig.~\ref{fig:main}\,(a). Therefore, in this intermediate regime, a rapid increasing of the MSD can be found. 

To further investigate the scaling, we expand Eq.~\eqref{eq:Cov} in time $t$ around $t=0$, focusing on short-time regime. 
Up to a leading order, the MSD shows linear term. 
\begin{align}
    \langle |\hat x(  t) - \hat x(0)|^2 \rangle =  \frac{\hbar(\nu_-+\nu_+)t}{2m\omega} +\mathcal{O}(t^2)
\end{align}
Unlike the Brownian motion of an underdamped classical particle, the ballistic regime does not appear at early times when the particle is subject to the quantum heat bath~$\mathcal D[\hat \rho_c]$.
When we set $\nu_-$ and $\nu_+$ as $0$, this leads to
\begin{align}
    \langle |\hat x(  t) - \hat x(0)|^2 \rangle  = \frac{\omega\hbar}{2m} t^2
-\frac{\omega^3\hbar}{24m} t^4
+\left(\frac{D_u \tau \omega^4}{36}
+\frac{\hbar\omega^5}{720m}\right) t^6
-\frac{D_u \omega^4}{168} t^7 
+ \mathcal{O}(t^8).
\label{eq:propiniMSD}\end{align}
Eq.~\eqref{eq:propiniMSD} shows that a contribution $t^6$ exists and suggests that $t^6$ scaling can be observed with high $D_u$ and $\tau$. Also, positive $t^7$ scaling is prohibited since the term has a negative sign. 
Note that setting $\nu_-=0$ and $\nu_+=0$ removes the early linear term ($(\nu_-+\nu_+)t\hbar /2m\omega$) in the above expression. As Eq.~\eqref{eq:propiniMSD} is expanded to $t=0$, Eq.~\eqref{eq:propiniMSD} does not explain the MSD at late times. 
From Eq.~\eqref{eq:propiniMSD}, a characteristic time~$\tau^*$ when MSD starts to increase rapidly can be derived
\begin{align}
    \tau^* \equiv \left[\frac{\omega\hbar}{2m}\right]^{\frac{1}{4}}\left[ \frac{D_u\tau \omega^4}{36} + \frac{\hbar \omega^5}{720 m} \right ]^{-\frac{1}{4}}
\label{eq:taustar}.\end{align}
The threshold time, $\tau^*$, is obtained by equating the $t^2$ and $t^6$ terms in Eq.~\eqref{eq:propiniMSD}. Eq.~\eqref{eq:taustar} predicts that increasing the noise intensity of the internal active state, $D_u$, leads to an earlier onset of the abrupt increase. A similar behavior is observed in the $\hbar \to 0$ limit.

The $t^6$ scaling can be understood directly from Eq.~\eqref{eq:OULangevin}, which provides an intuitive classical picture of the quantum dynamics. When $u$ is initially sampled from its steady state distribution [Eq.~\eqref{eq:inicond1}], $x_c(t)$ exhibits persistent motion and grows approximately linearly in time over short times. Consequently, the force exerted on the oscillator through the displaced trap center also grows linearly in time. Neglecting dissipation on this time scale, the quasiclassical Langevin equation gives
\begin{align}
    m\ddot{x} &\sim t, \label{eq:linear_t}
\end{align}
, which can be interpreted from the point of view of the jerky active matter~\cite{teVrugt2021Jerky,antonov2026jerky}, and, after integrating Eq.~\eqref{eq:linear_t} twice and squaring the displacement, the MSD of $x$ becomes 
\begin{align}
    |x(t)-x(0)|^2 &\sim t^6 
.\label{eq:physicst6}\end{align}
Therefore, the corresponding contribution to the MSD scales as $t^6$. The $t^6$ regime can thus be understood as a consequence of a persistently increasing force.

Although the $t^6$ scaling arises from the persistent motion of $x_c$, we do not regard it as unique signature of activity. Based on the above argument, similar scaling may be observed in a harmonically trapped system subject to deterministic driving with the corresponding temporal behavior. Recently, in the context of jerky active matter, it is reported that the $t^6$ scaling can also emerge from the interplay between activity and dry friction~\cite{antonov2026jerky}.

We note that Eq.~\eqref{eq:criterion_main} is not satisfied for the parameter choices used to illustrate $t^{6}$ and $t^7$ in the following figures 
Fig.~\ref{fig:t6activity} and Fig.~\ref{fig:t7t6activity}. However, since we choose small values of $\nu_-$ and $\nu_+$, the characteristic timescales associated with the Lindblad dynamics, $\nu_-^{-1}$ and $\nu_+^{-1}$, are sufficiently long that its effects remain limited over the time window shown in Fig.~\ref{fig:t6activity} and Fig.~\ref{fig:t7t6activity}.

In Fig.~\ref{fig:t6activity}, we plot MSDs of the system mimicking quantum active matter varying $D_u$. We choose small $\nu_-$ and $\nu_+$ and large $D_u$ to observe a $t^6$ scaling. 
As $D_u$ increases, the slope in the intermediate regime approaches to $t^6$. For the case $D_u = 10^6$, we performed a fit in the intermediate time regime using the ansatz $d\, t^{\alpha}$ where $d$ and $\alpha$ are real values.  
The resulting exponent is $\alpha = 5.9577...\pm 0.00296$, which is very close to $6$. For this $t^6$ scaling, small $\nu_-$ and $\nu_+$ (weak dissipation) are necessary. Therefore, the conditions for observing $t^6$ scaling and the absence of $t^7$ scaling are verified. Also, we find that Eq.~\eqref{eq:taustar}, shown as a vertical line, accurately predicts the onset of the abrupt increase.

The $t^6$ scaling was originally predicted in Ref.~\cite{antonov2025engineering} for the nondissipative case using time-dependent perturbation theory. In that work, the external driving was treated as a perturbation to the quantum harmonic oscillator, $\hat{V}_p(x_c(t))=m\omega^2 x_c^2(t)/2 -m\omega^2 \hat x x_c(t) $, so that
\begin{align}
    \hat{H}(x_c(t)) = \frac{\hat{p}^2}{2m} + \frac{1}{2}m\omega^2\hat{x}^2 + \hat{H}_p(x_c(t))
.\end{align}
Assuming the perturbation to be small, \citet{antonov2025engineering} showed that, at second order in perturbation theory, the leading contribution to the MSD scales as $t^6$ for small $t$ , while all lower-order terms vanish.
In contrast, in the present work we obtain an approximate expression for the MSD by considering the short-time limit. Eq.~\eqref{eq:propiniMSD} therefore supports the $t^6$ scaling from a different perspective, and the explicit coefficients further reveal the conditions under which the $t^6$ scaling can be observed.

Table~\ref{table1} summarizes the scaling behavior of the MSD in different time regimes. 
We distinguish between long and short persistence times $\tau$, and indicate how the MSD depends on time in the very early, early, intermediate, late, and long-time limit ($t \gg \tau$) regimes. 
The early time is mainly governed by $\omega$, $\nu_-$ and $\nu_+$. In the weak dissipation case ($\nu_-,\nu_+ \ll \omega, 1/\tau $), the duration of the linear regime decreases and ballistic behavior can be observed. 
In contrast,  when $1/\gamma$ is shorter than or comparable to both $1/\omega$ and $\tau $, the MSD of $\hat x$ converges to that of $x_c$ due to strong dissipation before the $t^6$ scaling regime can emerge. 
The intermediate regime are started to be affected by $\tau$ and $D$.
In particular, the intermediate regime may exhibit nontrivial power-law scaling, $t^\alpha$ with $\alpha \leq 6$, or oscillatory behavior depending on the dissipation strength. When $1/\gamma< 1/\omega $, oscillatory behavior vanished. High $D_u \tau$ increases $\alpha$ and $\alpha = 6$ can be observed with small $\nu_-$ and $\nu_+$.
For long persistence times, the late-time dynamics approaches a ballistic ($t^2$) regime before crossing over to diffusive ($t$) scaling, whereas for short $\tau$ the crossover structure is modified accordingly.

\subsection{
MSD with the initial condition $u(0)=0$ and
$t^7$ scaling}

\begin{figure}
\includegraphics[width=0.48\textwidth]{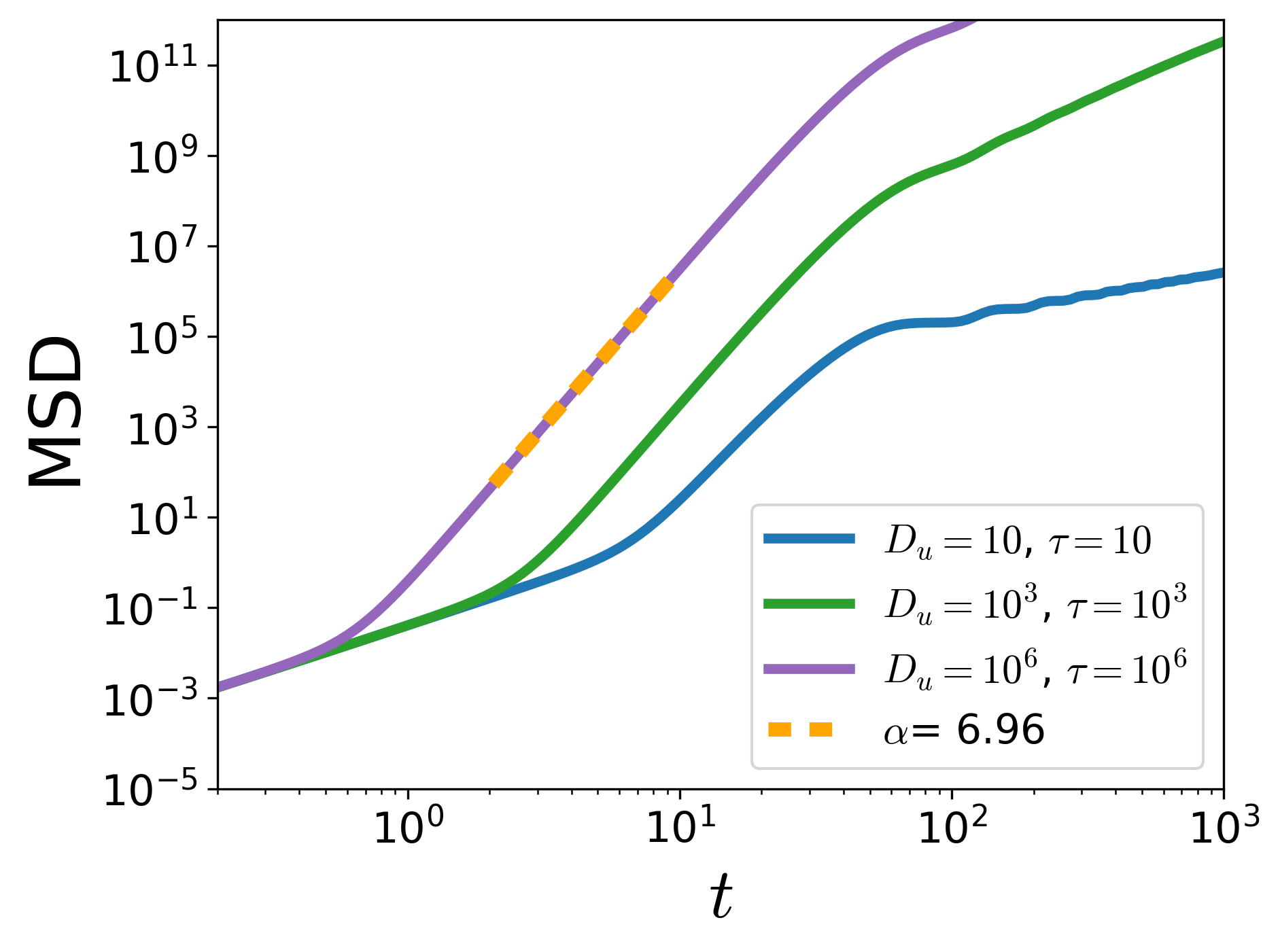}
\vskip -0.1in
\caption{ Plot of MSD versus time of the quantum active particle with different persistence time and noise intensity for the activity. 
We set $D_u$ and $\tau$ to be equal, they are varying from $10$ to $10^6$.
The blue solid line represents the case of long persistence time. 
The frequency of the quantum harmonic trap is chosen as $\omega=0.08$. The dissipation and pumping rates are $\nu_- = 10^{-4}$ and $\nu_+ = 10^{-8}$, respectively. We set the mass and Planck constant to unity.
}
\label{fig:t7t6activity}
\vskip -0.1in
\end{figure}

Now, we consider the same initial condition as in Refs.~\cite{antonov2025engineering,antonov2025modeling}.
The only difference is the distribution of $u$.
Explicitly, this initial condition is the joint distribution of ground state of the harmonic oscillator centered at $x = 0$,
while 
\begin{align}
    P(x_c, u) = \delta(x_c) \delta(u)
.\label{eq:delta_ini}\end{align}
This results in the MSD of $x_c$ shows the transient behavior from $t^3$ scaling at short times to $t$ at long times.
Explicitly, 
\begin{align}
    \langle |x_c(t) - x_c(0)|^2 \rangle =D_u \tau^2 [(-3 - e^{-((2 t)/\tau)} + 4 e^{-(t/\tau)}) \tau + 2 t]
.\label{eq:t3t1}\end{align} For the derivation, see Appendix~\ref{append:deriv}. By Taylor expanding Eq.~\eqref{eq:t3t1} with respect to time $t$ at $t=0$, one can check that $t^3$ is dominant at early times. This $t^3$ scaling of Eq.~\eqref{eq:t3t1} implies that the initial condition \eqref{eq:delta_ini} may lead to $\alpha >6$ scaling of MSD of $\hat x$.

We numerically evaluated Eq.~\eqref{eq:Cov} with the initial condition \eqref{eq:delta_ini} and plot the results in Fig.~\ref{fig:t7t6activity}. 
With a long persistence time and a large activity strength $D_u$, we observe a distinct scaling behavior characterized by a $t^{7}$ scaling. 
Specifically, the MSD initially grows as $t^{2}$ around $t \sim 10^{0}$, crosses over to a $t^{7}$ scaling regime near $t \sim 10^{1}$, and returns to a $t^{2}$ growth for $t \gtrsim 10^{2}$. 
In the very short-time regime, $t \sim 10^{-4}$, we observe a linear growth of the MSD induced by diffusivity. 
By contrast, for relatively smaller activity strength $D_u$ and shorter persistence times, we recover the previously reported $t^{6}$ scaling behavior.

As in Fig.~\ref{fig:t6activity}, Eq.~\eqref{eq:criterion_main} is not satisfied for the parameter choices used in Fig.~\ref{fig:t7t6activity}. Nevertheless, since $\nu_-$ and $\nu_+$ are chosen to be small, even the shorter characteristic timescale of the Lindblad dynamics, $\min{\nu_-^{-1},\nu_+^{-1}}$, is long compared with the time window shown in Fig.~\ref{fig:t7t6activity}. Thus, the effects of Eq.~\eqref{eq:Lindblad-dynamic} remain limited in Fig.~\ref{fig:t7t6activity}.

The analytical expression allows us to identify the conditions under which the $t^6$ and $t^7$ scaling regimes appear. Also, the reason why $t^8$ cannot be observed can be found from the following analytical expansion. In the limit of small pumping and dissipation rates, where a $t^6$ scaling was found~\cite{antonov2025engineering}, the expression is given by 
\begin{align}
\langle |\hat x( t) - \hat x(0)|^2 \rangle = \;&
\frac{\omega\hbar}{2m}\,  t^{2}
-\frac{\omega^{3}\hbar}{24m}\,  t^{4}
+\frac{\omega^{5}\hbar}{720m}\,  t^{6}+\frac{D_u\, \omega^{4}}{126}\,  t^{7} 
-\frac{\omega^{4}}{40320\, m\, \tau}
\left(140\, D_u\, m + \tau\, \omega^{3}\hbar\right)
 t^{8} + \mathcal O (t^9)
.\label{eq:analytic}\end{align} 
The above expression is derived in the small time limit, $\delta t\rightarrow 0$. Eq.~\eqref{eq:analytic} implies the existence of a $t^2$ slope, $t^6$ slope and $t^7$ slope. 
Interestingly, a term involving $D_u$ does not appear before the 7-th order term, and a term involving $1/\tau$ does not appear before the 8-th order term. In addition to that, the negative sign of $t^8$ term indicates that the increasing slope of $t^8$ in the MSD cannot be found in this limit. 
As the sign of the $t^4$ term is also negative, the $t^4$ slope is absent in this regime. Therefore, a large activity strength suppresses all contributions below seventh order in time and the long persistence time suppresses the eighth order term. This results in $t^7$ slope in the MSD plot of large $D_u$ and long $\tau$. Conversely, with small $D_u$ reveals $t^6$ slope as the small intensity suppressed $t^7$ term. It should be noted that Eq.~\eqref{eq:analytic} relies on the small-time approximation and the assumption of weak $\nu_-$ and $\nu_+$, and thus does not capture the full time dependence of the MSD. 

Following the reasoning of section IV A for the $t^6$ scaling, the $t^7$ scaling can also be explained using the physical intuition provided by the quasiclassical Langevin equation. For the initial condition $u(0)=0$, a stochastic kick is necessary for $x_c$ to move. This gives the short-time scaling of the MSD of $x_c$
\begin{align}
\left\langle [x_c(t)-x_c(0)]^2 \right\rangle \sim t^3
,\end{align}
which implies that the right-hand side of the quasiclassical Langevin equation is proportional to $t^{3/2}$.
Therefore, after double integration, one may find $ |x(t)-x(0)| \sim t^{7/2}$, which leads to the $t^7$ scaling of the MSD of $x$.

\begin{figure}[b]
    \centering
    \includegraphics[width=0.5\linewidth]{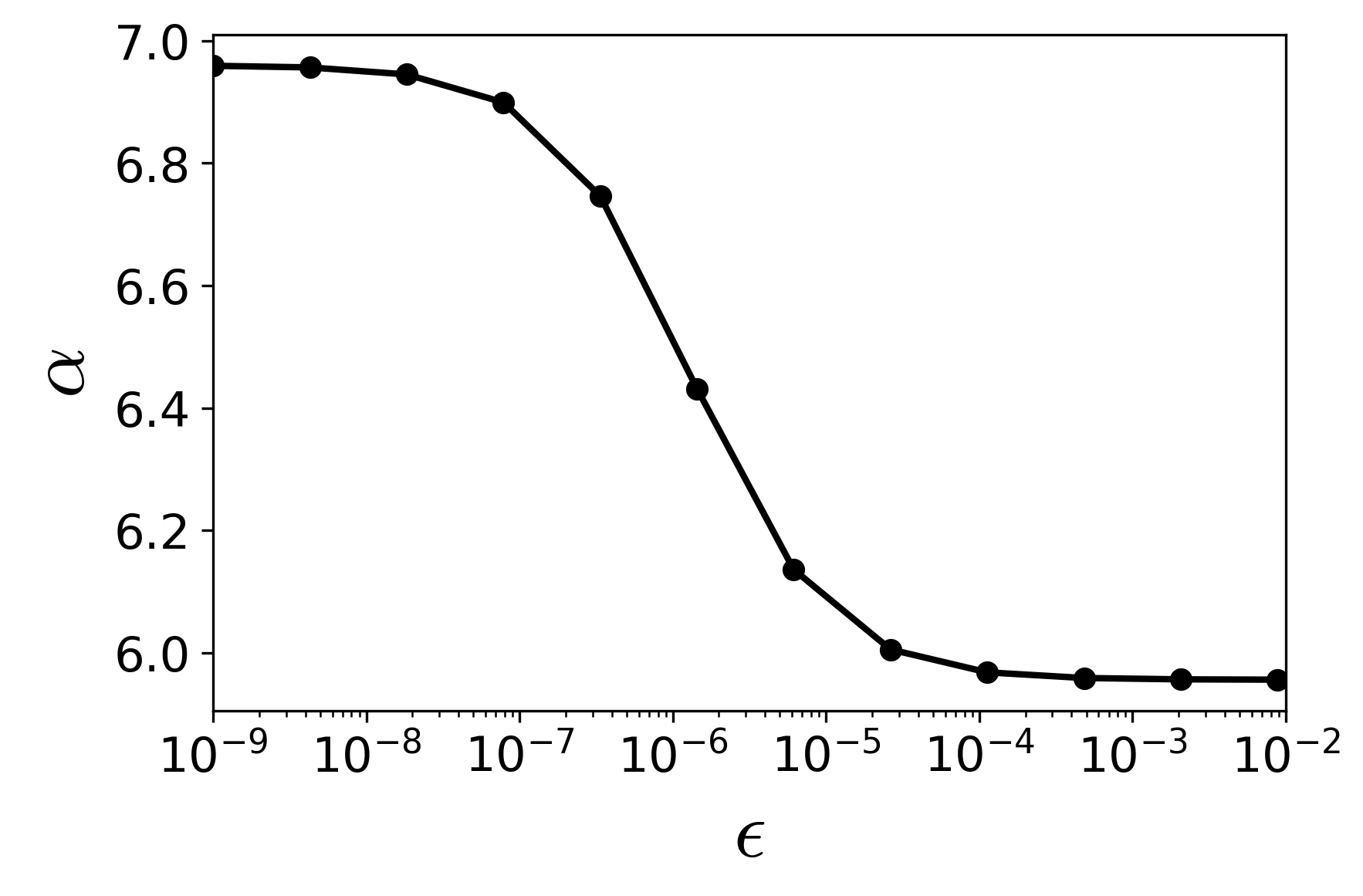}
    \caption{ 
    Plot of $\alpha$ versus $\epsilon$ where the variance of the initial distribution $u$ is given by $\epsilon D_u \tau$. $\alpha$ is estimated in the same time regime where $\alpha$ estimated in Fig.~\ref{fig:t7t6activity}. Except for the variance, all other parameters are set to the same values as those used for the $D_u=10^6$ case in Fig.~\ref{fig:t7t6activity}.
    }
    \label{fig:ealphaplot}
\end{figure}

Figs.~\ref{fig:t6activity} and ~\ref{fig:t7t6activity}, together with the analysis so far, show that $\alpha$ is sensitive to the distribution of $u$. 
Therefore, we examine how $\alpha$ changes as the distribution of $u$ varies. We plot $\alpha$, estimated in the same time regime used for estimating $\alpha$ in Fig.~\ref{fig:ealphaplot}, while varying the variance of the initial condition from $10^{-9}$ to $10^{-2}$ times its steady state value. Specifically, we introduce $\epsilon$ and the variance of the initial distribution of $u$ is set to $\epsilon D_u \tau$. 
In the semi-log plot, we find that $\alpha $ changes as a decreasing sigmoid function. 

\section{the effect of the initial quantum state}

\begin{figure}
\includegraphics[width=0.48\textwidth]{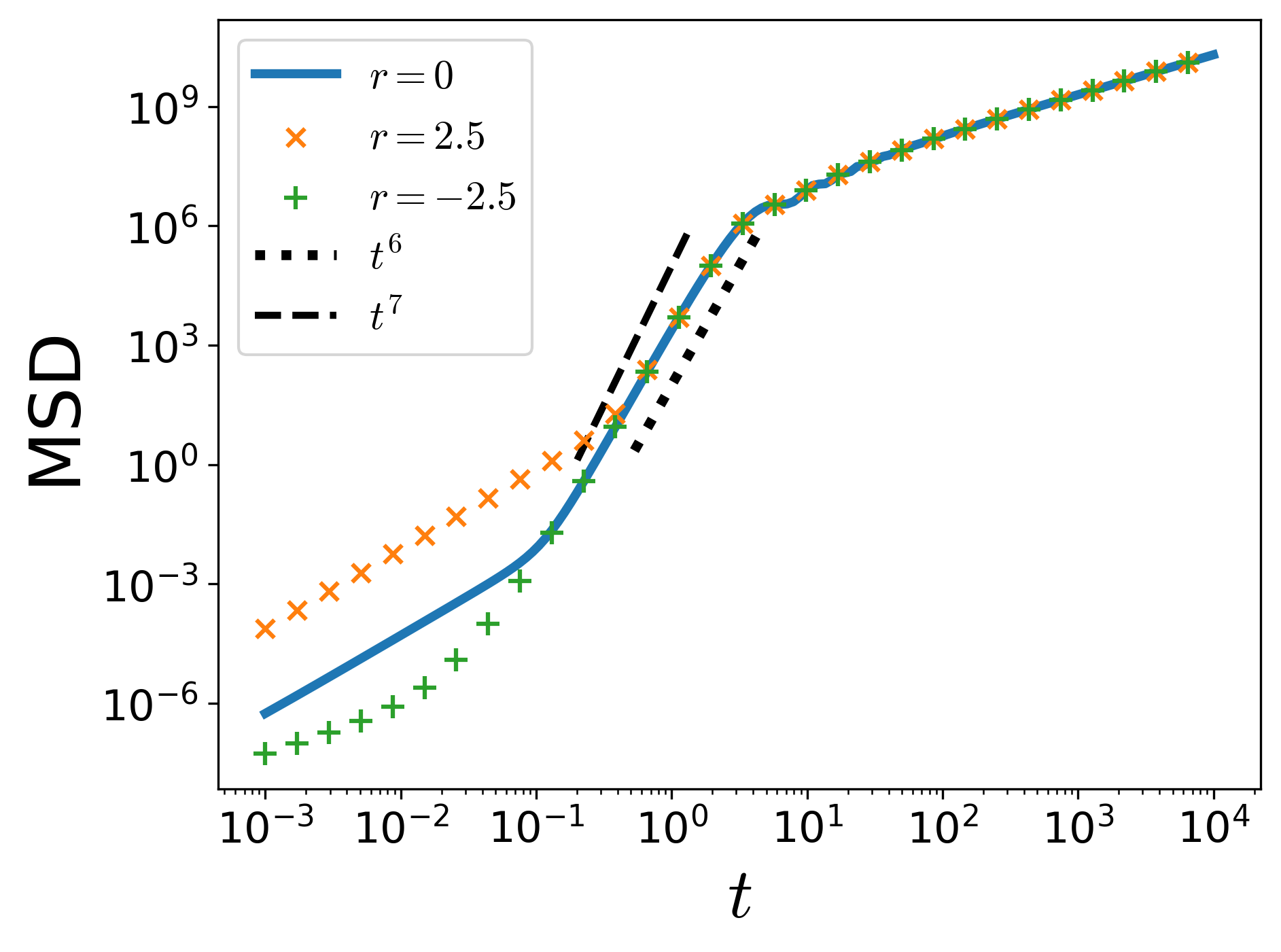}
\vskip -0.1in
\caption{ Plot of MSD [Eq.~\eqref{eq:Cov}] versus time of the quantum active particle with three initial conditions.
The initial distribution for classical variables are chosen as Eq.~\eqref{eq:inicond1}.
The orange `$\times$' marker shows the result of initially squeezed state~[Eq.~\eqref{eq:squeezedinistate}] with $r=2.5$ and the green `+' marker shows the results of initially squeezed state with $r=-2.5$. 
The solid line represents the unsqueezed case. We set $\omega=1$, $D_u =1000$ and $\tau=1000$. The other parameters are same as that is used for plotting Fig.~\ref{fig:t7t6activity}, for which the behavior scales as $t^6$.
}
\label{fig:squeez1}
\vskip -0.1in
\end{figure}

\begin{figure}
\includegraphics[width=0.48\textwidth]{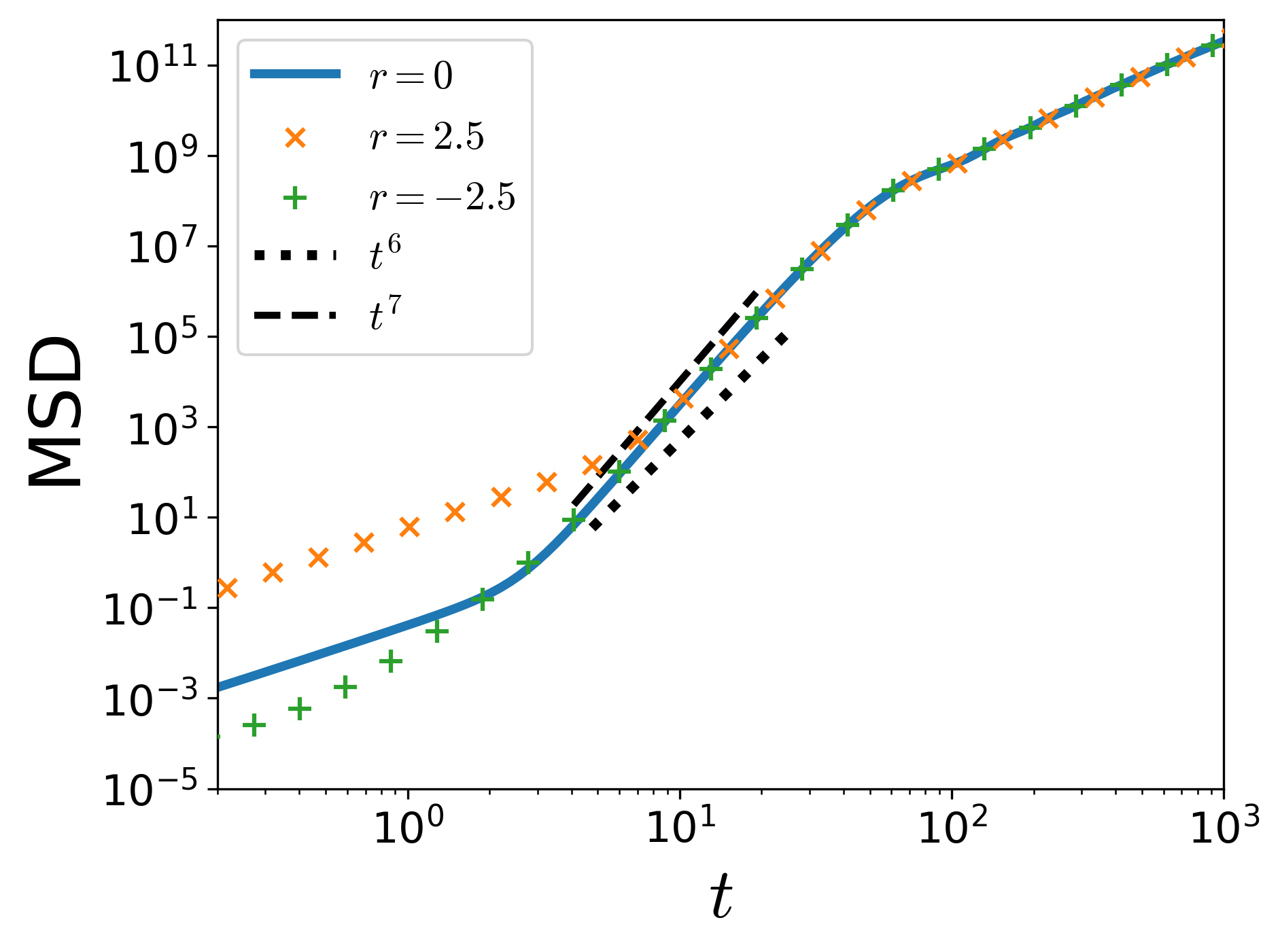}
\vskip -0.1in
\caption{ Plot of MSD~[Eq.~\eqref{eq:Cov}] versus time of the quantum active particle with three initial conditions. The initial distribution for classical variables are chosen as Eq.~\eqref{eq:delta_ini}.
The blue line represents the result for an initial ground state of a harmonic trap centered at $x_c=0$, which is the same result as in Fig.~\ref{fig:t7t6activity} ($D_u=1000 $ and $\tau =1000$). The orange `$\times$' symbol represent the result with a squeezed initial state~[Eq.~\eqref{eq:squeezedinistate}] of $r=2.5$. The green `+' symbol represents the result with squeezed initial state of $r=-2.5$. 
The frequency of the quantum harmonic trap is chosen as $\omega=0.08$. The dissipation and pumping rates are $\nu_- = 10^{-4}$ and $\nu_+ = 10^{-8}$, respectively. The persistence time~$\tau$ for the activity is 1000, and $D_u$ is also chosen as 1000. We set the mass and Planck constant to unity. 
}
\label{fig:squeez2}
\vskip -0.1in
\end{figure}

We have shown  that the $t^6$ and $t^7$ scalings highly depend on the initial distribution of the state variables of activity ($x_c, u$). One may wonder then how the initial quantum state changes the scalings. For this purpose, we consider a squeezed initial quantum state. Squeezed quantum states can have sharp distribution on position or momentum space while conserving the uncertainty relation. 

We computed the MSD for various initial conditions by varying the squeezing parameter $r$ and present a comparative plot in Figs.~\ref{fig:squeez1} and \ref{fig:squeez2}. 
Explicitly, the squeezed state can be expressed as 
\begin{align}
    |\psi_{\rm ini}\rangle =\hat {S}(r)|0\rangle 
\label{eq:squeezedinistate}\end{align} 
where $|0\rangle $ is the ground state of the harmonic oscillator at time $t=0$ and the squeezing operator is given by 
\begin{align}
    \hat{S}(r)
= \exp [\frac{r}{2}
(\hat{a}^2 - \hat{a}^{\dagger 2})].
\end{align}
Here, $r$ is a real value (for detailed information regarding squeezing operators see section 3.4.3 in Ref.~\cite{breuer2002theory}). 
Squeezing with a positive $r$ results in a position distribution narrowed by a factor of $e^{-2r}$ and a momentum space distribution broadened by a factor $e^{2r}$. The uncertainty principle thereby continues to hold. 

For positive values of $r$, the initial state exhibits a sharply localized position distribution accompanied by a broadened momentum distribution. Conversely, for negative values of $r$, the position distribution becomes broadened while the momentum distribution is sharply localized. In all cases, we observe that the $t^6$ and $t^{7}$ scalings of the MSD emerges robustly, indicating that this anomalous scaling behavior is insensitive to the initial quantum state. Also, we find that for $r>0$ the MSD grows faster than in the unsqueezed case ($r=0$), whereas for $r<0$ the growth of the MSD is comparatively slower. This behavior can be attributed to the fact that a broader momentum distribution corresponds to a larger initial kinetic energy, which in turn leads to a more rapid increase of the MSD.

\section{conclusion}

In this paper, we revisit the anomalous scaling behavior of the MSD in a system that mimics quantum active matter, and we demonstrated that the hybrid Wigner transformation provides a practical framework for analyzing quantum active matter that activeness is originated from classical dynamics. 
Extending previous analyses~\cite{antonov2025engineering,antonov2025modeling}, we evaluate the MSD without relying on the long-time approximation. While the underlying master equation remains valid beyond the small-$\gamma$ regime as in Ref.~\cite{antonov2025modeling}, our approach enables a systematic analysis of anomalous scaling beyond the small-$\gamma$ limit considered in Ref.~\cite{antonov2025engineering}. 
We also consider initial conditions for which the effects of activity appear already at $t=0$. In addition, we include the initial conditions used in previous studies~\cite{antonov2025engineering,antonov2025modeling}

Using a hybrid Wigner master equation, we derive an analytical expression for the MSD and evaluate it numerically. Compared with the trajectory-based approach of Ref.~\cite{antonov2025modeling}, which requires repeated simulations of the quantum dynamics for different realizations of $x_c$, the analytically evaluated MSD in our plots is free from statistical fluctuations associated with finite sample sizes.
The improved resolution of the MSD, together with the absence of the long-time approximation, allows for a more precise characterization of its scaling behavior.
With the initial condition for which the effects of activity appear already at $t=0$, a clearer $t^6$ scaling is observed. 
Furthermore, for the initial conditions employed in earlier works~\cite{antonov2025engineering,antonov2025modeling}, we find that the MSD can exhibit an even steeper scaling regime characterized by $t^7$ growth in certain parameter regimes, particularly with long persistence time $\tau$ of the active noise and in the limit of large diffusion coefficient $D_u$. 
In addition, the enhanced resolution enables a clearer resolution of oscillatory behavior in the intermediate-time regime.
Under the approximations of small pumping and dissipation rates, we further derive simplified analytical expressions for the MSD that provide insight into the origin of the $t^6$ and $t^7$ scaling behaviors. The obtained expressions and physical explanation from the quasi-Langevin equation with intuitions from classical physics support the aforementioned conditions under which $t^6$ and $t^7$ scaling emerges.

The exact solvability of the MSD for this model follows from master equation's linearity. For more general models, the hybrid Wigner master equation, together with the corresponding quasiclassical stochastic equation where available, may still provide a useful route for applying techniques developed in the classical active-matter field and classical intuition.

Finally, we examine the robustness of these anomalous scaling regimes with respect to the choice of the initial quantum state. We consider squeezed initial states and demonstrate that the exotic scaling behavior persists. These results highlight the stability of the anomalous scalings and provide a useful framework for analyzing nonequilibrium dynamics in quantum active matter.

Our analysis has been based on the adiabatic master equation, whose validity imposes a constraint on the trap motion that becomes important when $\nu_-$ and $\nu_+$ are large. Since the effects of the Lindblad dynamics are weak for small $\nu_-$ and $\nu_+$ over the relevant time window, it would be interesting to go beyond the adiabatic approximation and examine how nonadiabatic effects in the large-$\nu_\pm$ regime influence the mean-squared displacement.

\begin{acknowledgments}
M.t.V.\ is funded by the Deutsche Forschungsgemeinschaft (DFG, German Research Foundation) -- SFB 1551, Project-ID  464588647 and by the Carl-Zeiss-Stiftung (Wildcard programme). Funding for the position of Y.T.\ was provided by the DFG in the framework of TRR 146, Project-ID 233630050. This study contributes to research done in the Mainz Institute of Multiscale Modeling, M$^3$ODEL.
\end{acknowledgments}

\section*{Code availability}
We provide the Wolfram code~\cite{Tuchkov2026WolframCode} used to analytically handle the Wigner transformation, which may facilitate further analytical and numerical studies of quantum active matter. 

\bibliography{QAMnew}

@article{Javanainen1980Laser,
	author = {Javanainen, J. and Stenholm, S.},
	date = {1980/03/01},
	doi = {10.1007/BF00886180},
	id = {Javanainen1980},
	isbn = {1432-0630},
	journal = {Applied physics},
	number = {3},
	pages = {283--291},
	title = {Laser cooling of trapped particles I: The heavy particle limit},
	url = {https://doi.org/10.1007/BF00886180},
	volume = {21},
	year = {1980}}

@article{Schutz2013Cooling,
  title = {Cooling of atomic ensembles in optical cavities: Semiclassical limit},
  author = {Sch\"utz, Stefan and Habibian, Hessam and Morigi, Giovanna},
  journal = {Phys. Rev. A},
  volume = {88},
  issue = {3},
  pages = {033427},
  numpages = {13},
  year = {2013},
  month = {Sep},
  publisher = {American Physical Society},
  doi = {10.1103/PhysRevA.88.033427},
  url = {https://link.aps.org/doi/10.1103/PhysRevA.88.033427}
}

@article{antonov2026jerky,
  title={Jerky Motion of Active Granular Particles},
  author={Antonov, Alexander P and Musacchio, Marco and L{\"o}wen, Hartmut and Caprini, Lorenzo},
  journal={arXiv preprint arXiv:2608.24689},
  year={2026}
}

@article{teVrugt2021Jerky,
doi = {10.1088/1367-2630/abfa61},
url = {https://doi.org/10.1088/1367-2630/abfa61},
year = {2021},
month = {jun},
publisher = {IOP Publishing},
volume = {23},
number = {6},
pages = {063023},
author = {te Vrugt, Michael and Jeggle, Julian and Wittkowski, Raphael},
title = {Jerky active matter: a phase field crystal model with translational and orientational memory},
journal = {New Journal of Physics}
}

@article{Scholz2018Inertial,
	author = {Scholz, Christian and Jahanshahi, Soudeh and Ldov, Anton and L{\"o}wen, Hartmut},
	date = {2018/12/04},
	doi = {10.1038/s41467-018-07596-x},
	id = {Scholz2018},
	isbn = {2041-1723},
	journal = {Nature Communications},
	number = {1},
	pages = {5156},
	title = {Inertial delay of self-propelled particles},
	url = {https://doi.org/10.1038/s41467-018-07596-x},
	volume = {9},
	year = {2018}}

@article{Gipouloux2026Active,
       author = {{Gipouloux}, Jeanne and {Brunelli}, Matteo and {Cugliandolo}, Leticia and {Fazio}, Rosario and {Schir{\`o}}, Marco},
        title = "{Active Quantum Particles from Engineered Dissipation}",
         year = 2026,
        month = mar,
        journal = {arXiv:2603.19094},
          doi = {10.48550/arXiv.2603.19094},
 primaryClass = {quant-ph},
       adsurl = {https://ui.adsabs.harvard.edu/abs/2026arXiv260319094G}
}

@article{
Vijay2007Long,
author = {Vijay Narayan  and Sriram Ramaswamy  and Narayanan Menon },
title = {Long-Lived Giant Number Fluctuations in a Swarming Granular Nematic},
journal = {Science},
volume = {317},
number = {5834},
pages = {105-108},
year = {2007},
doi = {10.1126/science.1140414},
URL = {https://www.science.org/doi/abs/10.1126/science.1140414},
eprint = {https://www.science.org/doi/pdf/10.1126/science.1140414}}

@article{Diosi2023Hybrid,
  title = {Hybrid completely positive Markovian quantum-classical dynamics},
  author = {Di\'osi, Lajos},
  journal = {Phys. Rev. A},
  volume = {107},
  issue = {6},
  pages = {062206},
  numpages = {10},
  year = {2023},
  month = {Jun},
  publisher = {American Physical Society},
  doi = {10.1103/PhysRevA.107.062206},
  url = {https://link.aps.org/doi/10.1103/PhysRevA.107.062206}
}

@article{Schuettler2025Active,
  title = {Active particles in moving traps: Minimum work protocols and information efficiency of work extraction},
  author = {Sch\"uttler, Janik and Garcia-Millan, Rosalba and Cates, Michael E. and Loos, Sarah A. M.},
  journal = {Phys. Rev. E},
  volume = {112},
  issue = {2},
  pages = {024119},
  numpages = {21},
  year = {2025},
  month = {Aug},
  publisher = {American Physical Society},
  doi = {10.1103/4q4f-1dpx},
  url = {https://link.aps.org/doi/10.1103/4q4f-1dpx}
}

@article{kalz2024field,
  title={Field theory of active chiral hard disks: a first-principles approach to steric interactions},
  doi={10.1088/1751-8121/ad5089},
  author={Kalz, Erik and Sharma, Abhinav and Metzler, Ralf},
  journal={J. Phys. A: Math. Theor.},
  volume={57},
  number={26},
  pages={265002},
  year={2024},
  publisher={IOP Publishing}
}

@article{lemaitre2023non,
  title={Non-{G}aussian displacement distributions in models of heterogeneous active particle dynamics},
  author={Lemaitre, Elisabeth and Sokolov, Igor M and Metzler, Ralf and Chechkin, Aleksei V},
  doi={10.1088/1367-2630/acb005},
  journal={New J. Phys.},
  volume={25},
  number={1},
  pages={013010},
  year={2023},
  publisher={IOP Publishing}
}

@article{grossmann2024non,
  title={Non-{G}aussian displacements in active transport on a carpet of motile cells},
  author={Gro{\ss}mann, Robert and Bort, Lara S and Moldenhawer, Ted and Stange, Maike and Panah, Setareh Sharifi and Metzler, Ralf and Beta, Carsten},
  doi={10.1103/PhysRevLett.132.088301},
  journal={Phys. Rev. Lett.},
  volume={132},
  number={8},
  pages={088301},
  year={2024},
  publisher={APS}
}

@article{Szamel2014Self,
  title={Self-propelled particle in an external potential: Existence of an effective temperature},
  author={Szamel, Grzegorz},
  journal={Phys. Rev. E},
doi = {10.1103/PhysRevE.90.012111},
  volume={90},
  number={1},
  pages={012111},
  year={2014},
  publisher={APS}
}

@book{Rivas2012open,
  title={Open quantum systems},
  author={Rivas, Angel and Huelga, Susana F},
  volume={10},
  year={2012},
  publisher={Springer}
}

@misc{Tuchkov2026WolframCode,
  author       = {Tuchkov, Y. and Lee, S.},
  title        = {Symbolic Wigner Transformation (Version 1.0.0)},
  year         = {2026},
  journal      = {Zenodo},
  publisher    = {Zenodo},
  doi          = {https://doi.org/10.5281/zenodo.19693441},
  note          = {https://doi.org/10.5281/zenodo.19693441}
}

@article{Massignan2015Quantum,
  title = {Quantum Brownian motion with inhomogeneous damping and diffusion},
  author = {Massignan, Pietro and Lampo, Aniello and Wehr, Jan and Lewenstein, Maciej},
  journal = {Phys. Rev. A},
  volume = {91},
  issue = {3},
  pages = {033627},
  numpages = {20},
  year = {2015},
  month = {Mar},
  publisher = {American Physical Society},
  doi = {10.1103/PhysRevA.91.033627},
  url = {https://link.aps.org/doi/10.1103/PhysRevA.91.033627}
}

@misc{Mathematica,
  author = {Wolfram Research{,} Inc.},
  title = {Mathematica, {V}ersion 14.2},
  url = {https://www.wolfram.com/mathematica},
  note = {Champaign, IL, 2025}
}

@article{adachi2022activity,
  author = {Adachi, Kyosuke and Takasan, Kazuaki and Kawaguchi, Kyogo},
  doi = {10.1103/physrevresearch.4.013194},
  journal = {Phys. Rev. Res.},
  number = {1},
  pages = {013194},
  publisher = {APS},
  title = {Activity-induced phase transition in a quantum many-body system},
  volume = {4},
  year = {2022}
}

@article{albash2012quantum,
  author = {Albash, Tameem and Boixo, Sergio and Lidar, Daniel A and Zanardi, Paolo},
  doi = {10.1088/1367-2630/14/12/123016},
  journal = {New J. Phys.},
  number = {12},
  pages = {123016},
  publisher = {IOP Publishing},
  title = {Quantum adiabatic Markovian master equations},
  volume = {14},
  year = {2012}
}

@article{antonov2024inertial,
  author = {Antonov, Alexander P and Caprini, Lorenzo and Ldov, Anton and Scholz, Christian and L{\"o}wen, Hartmut},
  doi = {10.1103/physrevlett.133.198301},
  journal = {Phys. Rev. Lett.},
  number = {19},
  pages = {198301},
  publisher = {APS},
  title = {Inertial active matter with Coulomb friction},
  volume = {133},
  year = {2024}
}

@article{antonov2025engineering,
  author = {Antonov, Alexander P and Zheng, Yuanjian and Liebchen, Benno and L{\"o}wen, Hartmut},
  doi = {10.1103/z3gm-32jn},
  journal = {Phys. Rev. Res.},
  number = {3},
  pages = {033008},
  publisher = {APS},
  title = {Engineering active motion in quantum matter},
  volume = {7},
  year = {2025}
}

@article{antonov2025modeling,
  title = {Modeling dissipation in quantum active matter},
  author = {Antonov, Alexander P. and Lee, Sangyun and Liebchen, Benno and L\"owen, Hartmut and Melles, Jannis and Morigi, Giovanna and Tuchkov, Yehor and te Vrugt, Michael},
  journal = {Phys. Rev. A},
  volume = {114},
  issue = {1},
  pages = {012201},
  numpages = {14},
  year = {2026},
  month = {Jul},
  publisher = {American Physical Society},
  doi = {10.1103/4gyc-3pht},
  url = {https://link.aps.org/doi/10.1103/4gyc-3pht}
}

@article{aranson2022bacterial,
  author = {Aranson, Igor S},
  doi = {10.1088/1361-6633/ac723d},
  journal = {Rep. Prog. Phys.},
  number = {7},
  pages = {076601},
  publisher = {IOP Publishing},
  title = {Bacterial active matter},
  volume = {85},
  year = {2022}
}

@article{baconnier2022selective,
  author = {Baconnier, Paul and Shohat, Dor and L{\'o}pez, C Hern{\'a}ndez and Coulais, Corentin and D{\'e}mery, Vincent and D{\"u}ring, Gustavo and Dauchot, Olivier},
  doi = {10.1038/s41567-022-01704-x},
  journal = {Nat. Phys.},
  number = {10},
  pages = {1234},
  publisher = {Nature Publishing Group UK London},
  title = {Selective and collective actuation in active solids},
  volume = {18},
  year = {2022}
}

@article{Bechinger2016Active,
  author = {Bechinger, Clemens and Di Leonardo, Roberto and L\"owen, Hartmut and Reichhardt, Charles and Volpe, Giorgio and Volpe, Giovanni},
  doi = {10.1103/revmodphys.88.045006},
  issue = {4},
  journal = {Rev. Mod. Phys.},
  month = {Nov},
  numpages = {50},
  pages = {045006},
  publisher = {American Physical Society},
  title = {Active particles in complex and crowded environments},
  volume = {88},
  year = {2016}
}

@book{breuer2002theory,
  author = {Breuer, Heinz-Peter and Petruccione, Francesco},
  doi = {10.1093/acprof:oso/9780199213900.001.0001},
  publisher = {OUP Oxford},
  title = {The theory of open quantum systems},
  year = {2007}
}

@article{caprini2024emergent,
  author = {Caprini, Lorenzo and Ldov, Anton and Gupta, Rahul Kumar and Ellenberg, Hendrik and Wittmann, Ren{\'e} and L{\"o}wen, Hartmut and Scholz, Christian},
  doi = {10.1038/s42005-024-01540-w},
  journal = {Commun. Phys.},
  number = {1},
  pages = {52},
  publisher = {Nature Publishing Group UK London},
  title = {Emergent memory from tapping collisions in active granular matter},
  volume = {7},
  year = {2024}
}

@article{cavagna2014bird,
  author = {Cavagna, Andrea and Giardina, Irene},
  doi = {10.1146/annurev-conmatphys-031113-133834},
  journal = {Annu. Rev. Condens. Matter Phys.},
  number = {1},
  pages = {183},
  publisher = {Annual Reviews},
  title = {Bird flocks as condensed matter},
  volume = {5},
  year = {2014}
}

@article{Dann2018Time,
  author = {Dann, Roie and Levy, Amikam and Kosloff, Ronnie},
  doi = {10.1103/physreva.98.052129},
  issue = {5},
  journal = {Phys. Rev. A},
  month = {Nov},
  numpages = {14},
  pages = {052129},
  publisher = {American Physical Society},
  title = {Time-dependent Markovian quantum master equation},
  volume = {98},
  year = {2018}
}

@article{dekker1977quantization,
  author = {Dekker, H},
  doi = {10.1103/physreva.16.2126},
  journal = {Phys. Rev. A},
  number = {5},
  pages = {2126},
  publisher = {APS},
  title = {Quantization of the linearly damped harmonic oscillator},
  volume = {16},
  year = {1977}
}

@book{gardiner2004quantum,
  author = {Gardiner, Crispin and Zoller, Peter},
  publisher = {Springer Science \& Business Media},
  title = {Quantum noise: a handbook of Markovian and {non-Markovian} quantum stochastic methods with applications to quantum optics},
  year = {2004}
}

@book{gardiner2009stochastic,
  author = {Gardiner, Crispin},
  doi = {10.1007/978-3-662-02377-8},
  publisher = {Springer Berlin Heidelberg},
  title = {Stochastic methods},
  volume = {4},
  year = {1983}
}

@article{BurgardtEtAl2026,
  title={Quantum-enabled active matter at the atomic scale},
  author={Burgardt, Sabrina and Fe{\ss}, Julian and Guthmann, Alexander and Hiebel, Silvia and Mukhopadhyay, Aritra K and Lee, Sangyun and te Vrugt, Michael and Liebchen, Benno and L{\"o}wen, Hartmut and Wittkowski, Raphael and Widera, Artur},
  journal={arXiv:2606.24615},
  year={2026}
}

@article{ghosh2021enzymes,
  author = {Ghosh, Subhadip and Somasundar, Ambika and Sen, Ayusman},
  doi = {10.1146/annurev-conmatphys-061020-053036},
  journal = {Annu. Rev. Condens. Matter Phys.},
  number = {1},
  pages = {177},
  publisher = {Annual Reviews},
  title = {Enzymes as active matter},
  volume = {12},
  year = {2020}
}

@article{gompper20202020,
  author = {Gompper, Gerhard and Winkler, Roland G and Speck, Thomas and Solon, Alexandre and Nardini, Cesare and Peruani, Fernando and L{\"o}wen, Hartmut and Golestanian, Ramin and Kaupp, U Benjamin and Alvarez, Luis and others},
  doi = {10.1088/1361-648x/ab6348},
  journal = {J. Phys.: Condens. Matter},
  number = {19},
  pages = {193001},
  publisher = {IoP Publishing},
  title = {The 2020 motile active matter roadmap},
  volume = {32},
  year = {2020}
}

@incollection{groenewold1946principles,
  author = {Groenewold, Hilbrand Johannes},
  booktitle = {On the principles of elementary quantum mechanics},
  doi = {10.1016/s0031-8914(46)80059-4},
  pages = {405},
  publisher = {Springer},
  title = {On the principles of elementary quantum mechanics},
  volume = {12},
  year = {1946}
}

@article{jee2018catalytic,
  author = {Jee, Ah-Young and Cho, Yoon-Kyoung and Granick, Steve and Tlusty, Tsvi},
  doi = {10.1073/pnas.1814180115},
  journal = {Proc. Nat. Acad. Sci.},
  number = {46},
  pages = {E10812--E10821},
  publisher = {National Academy of Sciences},
  title = {Catalytic enzymes are active matter},
  volume = {115},
  year = {2018}
}

@article{khasseh2025active,
  author = {Khasseh, Reyhaneh and Wald, Sascha and Moessner, Roderich and Weber, Christoph A and Heyl, Markus},
  doi = {10.48550/arxiv.2308.01603},
  journal = {Phys. Rev. Lett.},
  number = {24},
  pages = {248302},
  publisher = {APS},
  title = {Active quantum flocks},
  volume = {135},
  year = {2023}
}

@article{Lee2020Finite,
  author = {Lee, Sangyun and Ha, Meesoon and Park, Jong-Min and Jeong, Hawoong},
  doi = {10.1103/physreve.101.022127},
  issue = {2},
  journal = {Phys. Rev. E},
  month = {Feb},
  numpages = {8},
  pages = {022127},
  publisher = {American Physical Society},
  title = {Finite-time quantum Otto engine: Surpassing the quasistatic efficiency due to friction},
  volume = {101},
  year = {2020}
}

@article{Lee2021Quantumness,
  author = {Lee, Sangyun and Ha, Meesoon and Jeong, Hawoong},
  doi = {10.1103/physreve.103.022136},
  issue = {2},
  journal = {Phys. Rev. E},
  month = {Feb},
  numpages = {12},
  pages = {022136},
  publisher = {American Physical Society},
  title = {Quantumness and thermodynamic uncertainty relation of the finite-time Otto cycle},
  volume = {103},
  year = {2021}
}

@article{liebchen2018synthetic,
  author = {Liebchen, Benno and Lowen, Hartmut},
  doi = {10.1021/acs.accounts.8b00215},
  journal = {Acc. Chem. Res.},
  number = {12},
  pages = {2982},
  publisher = {ACS Publications},
  title = {Synthetic chemotaxis and collective behavior in active matter},
  volume = {51},
  year = {2018}
}

@article{marchetti2013hydrodynamics,
  author = {Marchetti, M Cristina and Joanny, Jean-Fran{\c{c}}ois and Ramaswamy, Sriram and Liverpool, Tanniemola B and Prost, Jacques and Rao, Madan and Simha, R Aditi},
  doi = {10.1103/revmodphys.85.1143},
  journal = {Rev. Mod. Phys.},
  number = {3},
  pages = {1143},
  publisher = {APS},
  title = {Hydrodynamics of soft active matter},
  volume = {85},
  year = {2013}
}

@article{nadolny2025nonreciprocal,
  author = {Nadolny, Tobias and Bruder, Christoph and Brunelli, Matteo},
  doi = {10.1103/physrevx.15.011010},
  journal = {Phys. Rev. X},
  number = {1},
  pages = {011010},
  publisher = {APS},
  title = {Nonreciprocal synchronization of active quantum spins},
  volume = {15},
  year = {2025}
}

@article{penner2025heat,
  author = {Penner, Alexander-Georg and Viotti, Ludmila and Fazio, Rosario and Arrachea, Liliana and von Oppen, Felix},
  doi = {10.17169/refubium-50822},
  journal = {Phys. Rev. B},
  number = {18},
  pages = {L180303},
  publisher = {APS},
  title = {Heat-to-motion conversion for quantum active matter},
  volume = {112},
  year = {2025}
}

@mastersthesis{Qirezi2009,
  address = {London, United Kingdom},
  author = {Qirezi, Fatmir},
  school = {Queen Mary University of London},
  title = {From Stochastic to Deterministic Langevin Equation},
  type = {M.Sc. thesis},
  year = {2009}
}

@article{ramaswamy2010mechanics,
  author = {Ramaswamy, Sriram},
  doi = {10.1146/annurev-conmatphys-070909-104101},
  journal = {Annu. Rev. Condens. Matter Phys.},
  number = {1},
  pages = {323},
  publisher = {Annual Reviews},
  title = {The mechanics and statistics of active matter},
  volume = {1},
  year = {2010}
}

@article{ramaswamy2017active,
  author = {Ramaswamy, Sriram},
  doi = {10.1103/revmodphys.85.1143},
  journal = {J. Stat. Mech.},
  number = {5},
  pages = {1143},
  publisher = {IOP Publishing and SISSA},
  title = {Active matter},
  volume = {85},
  year = {2013}
}

@article{sastry1998signatures,
  author = {Sastry, Srikanth and Debenedetti, Pablo G and Stillinger, Frank H},
  doi = {10.1038/31189},
  journal = {Nature},
  number = {6685},
  pages = {554},
  publisher = {Nature Publishing Group UK London},
  title = {Signatures of distinct dynamical regimes in the energy landscape of a glass-forming liquid},
  volume = {393},
  year = {1998}
}

@article{schmidt2019light,
  author = {Schmidt, Falko and Liebchen, Benno and L{\"o}wen, Hartmut and Volpe, Giovanni},
  doi = {10.1063/1.5079861},
  journal = {J. Chem. Phys.},
  number = {9},
  pages = {094905},
  publisher = {AIP Publishing},
  title = {Light-controlled assembly of active colloidal molecules},
  volume = {150},
  year = {2019}
}

@article{soto2014self,
  author = {Soto, Rodrigo and Golestanian, Ramin},
  doi = {10.1103/physrevlett.112.068301},
  journal = {Phys. Rev. Lett.},
  number = {6},
  pages = {068301},
  publisher = {APS},
  title = {Self-assembly of catalytically active colloidal molecules: tailoring activity through surface chemistry},
  volume = {112},
  year = {2014}
}

@article{takasan2024activity,
  author = {Takasan, Kazuaki and Adachi, Kyosuke and Kawaguchi, Kyogo},
  doi = {10.1103/physrevresearch.6.023096},
  journal = {Phys. Rev. Res.},
  number = {2},
  pages = {023096},
  publisher = {APS},
  title = {Activity-induced ferromagnetism in one-dimensional quantum many-body systems},
  volume = {6},
  year = {2024}
}

@article{toner1995long,
  author = {Toner, John and Tu, Yuhai},
  journal = {Phys. Rev. Lett.},
  doi={10.1103/PhysRevLett.75.4326},
  number = {23},
  pages = {4326},
  publisher = {APS},
  title = {Long-range order in a two-dimensional dynamical {XY} model: how birds fly together},
  volume = {75},
  year = {1995}
}

@book{van1992stochastic,
  author = {Van Kampen, Nicolaas Godfried},
  doi = {10.1016/b978-0-444-52965-7.x5000-4},
  publisher = {Elsevier},
  title = {Stochastic processes in physics and chemistry},
  volume = {1},
  year = {2007}
}

@article{vicsek1995novel,
  author = {Vicsek, Tam{\'a}s and Czir{\'o}k, Andr{\'a}s and Ben-Jacob, Eshel and Cohen, Inon and Shochet, Ofer},
  doi = {10.1103/physrevlett.75.1226},
  journal = {Phys. Rev. Lett.},
  number = {6},
  pages = {1226},
  publisher = {APS},
  title = {Novel type of phase transition in a system of self-driven particles},
  volume = {75},
  year = {1995}
}

@article{vrugt2025exactly,
  title = {Colloquium: What do we mean by `active matter'?},
  author = {te Vrugt, Michael and Liebchen, Benno and Cates, Michael E.},
  journal = {Rev. Mod. Phys.},
  volume = {98},
  issue = {3},
  pages = {031001},
  numpages = {30},
  year = {2026},
  month = {Jul},
  publisher = {American Physical Society},
  doi = {10.1103/wd4f-q7kv},
  url = {https://link.aps.org/doi/10.1103/wd4f-q7kv}
}

@article{walther2008janus,
  author = {Walther, Andreas and M{\"u}ller, Axel HE},
  doi = {10.1039/b718131k},
  journal = {Soft Matter},
  number = {4},
  pages = {663},
  publisher = {Royal Society of Chemistry},
  title = {Janus particles},
  volume = {4},
  year = {2008}
}

@article{Wigner1932,
  author = {Wigner, E},
  doi = {10.1103/physrev.40.749},
  journal = {Phys. Rev.},
  pages = {749},
  title = {On the Quantum Correction for Thermodynamic Equilibrium},
  volume = {40},
  year = {1932}
}

\appendix
\onecolumngrid

\section{Equivalence between Wigner and operator expectation value}
\label{append:onetimeequiv} 
Consider the phase-space integral
\begin{align}
\int dx\,dp \, A(x,p) W(x,p)
&= \frac{2}{\pi \hbar}
\int dx\,dp\,dz\,dz'\,
e^{-\frac{2 i p}{\hbar}(z+z')}
\left\langle x+z \middle| \hat A \middle| x-z \right\rangle
\left\langle x+z' \middle| \hat\rho \middle| x-z' \right\rangle \nonumber \\
&= 2
\int dx\,dz\,dz'\,
\delta(z+z')
\left\langle x+z \middle| \hat A \middle| x-z \right\rangle
\left\langle x-z \middle| \hat\rho \middle| x+z \right\rangle \nonumber \\
&= 2
\int dx\,dz\,
\left\langle x+z \middle| \hat A \middle| x-z \right\rangle
\left\langle x-z \middle| \hat\rho \middle| x+z \right\rangle .
\end{align}
Here, performing the $p$ integration yields a delta function. 

By introducing new variables $s = x+z$ and $ s' = x-z $, the integral becomes
\begin{align}
\int dx\,dp \, A(x,p) W(x,p)
=&\int ds\,ds'\,
\left\langle s \middle| \hat A \middle| s' \right\rangle
\left\langle s' \middle| \hat\rho \middle| s \right\rangle\\ 
=& \mathrm{tr}\{ \hat A \hat\rho \}.
\label{eq:o_t_eq}\end{align}
Therefore, the average in the Wigner representation is same as the average with the corresponding quantum operator and the density matrix.

In the case of a two-time function, two expectations,
\begin{align}
    {\rm tr} \{ \hat A(t) \hat B(0) \hat \rho \} 
\end{align}
and
\begin{align}
    \int dx dp W[\hat A(t)](x,p) W[\hat B(0)](x,p)W(x,p)
,\end{align}
are not equivalent in general. Here, $W[\hat A]$ is defined as 
\begin{align}
    W[\hat A ] (x,y)
    \equiv \frac{1}{2\pi\hbar}
\iint dy d p_y\;
e^{\frac{i}{\hbar}(p_y x- y p)}\,
\chi[\hat\rho](y, p_y).
\end{align}
However, if $\hat A(t)$ and $\hat B(0)$ are linear to $\hat x$ and $\hat p$, then the two expectations 
\begin{align}
    \frac{1}{2} {\rm tr} \{ \hat A(t) \hat B(0) \hat \rho + \hat B(0) \hat A(t) \hat \rho\} \text{ and } 
    \int dx dp W[A(t)](x,p) W[B(0)](x,p)W(x,p)
\end{align}
are equivalent. 
In Appendix~\ref{append:HeisenClosed}, we explain the conditions when the evolution of moments is closed under the Lindblad master equation and can be a linear combination of basis operators.
 proof starts from Eq.~\eqref{eq:o_t_eq}:
\begin{align}
    {\rm tr}\left\{ (\hat A(t)\hat B(0) + \hat B(0)\hat A(t) ) \hat\rho\right\}  = \iint dxdp W[(\hat A(t)\hat B(0) + \hat B(0)\hat A(t) )] W(x,p,0)
\end{align}
The Wigner transformation of the symmetric operator is given by
\begin{align}
W\left[
\hat A(t)\hat B(0) + \hat B(0)\hat A(t)
\right] =& W [
\hat A(t)]\star  W[\hat B(0) ] + W [\hat B(0)]\star W[\hat A(t) ]
\end{align}
When $\hat A$ and $\hat B$ are linear combinations of $\hat x$ and $\hat p$, the above equality can be rearranged as 
\begin{align}
W\left[
\hat A(t)\hat B(0) + \hat B(0)\hat A(t)
\right]&= 2\, W[\hat A(t)]\, W[\hat B(0)] \nonumber \\
&\quad
+ \frac{i}{2} W[\hat A(t)]
\left(
\overleftarrow{\partial}_{x}
\overrightarrow{\partial}_{p}
-
\overleftarrow{\partial}_{p}
\overrightarrow{\partial}_{x}
\right)
W[\hat B(0)] \nonumber \\
&\quad
- \frac{i}{2} W[\hat B(0)]
\left(
\overleftarrow{\partial}_{x}
\overrightarrow{\partial}_{p}
-
\overleftarrow{\partial}_{x}
\overrightarrow{\partial}_{q}
\right)
W[\hat A(t)] \nonumber \\
&= 2 W[\hat A(t)]\, W[\hat B(0)] .
\end{align}
Therefore, the symmetrized two-time correlation function and the corresponding two-time correlation function in the Wigner representation are equivalent:
\begin{align}
\frac{1}{2}
\Big(
\mathrm{Tr}\{\hat A(t)\hat B(0)\hat \rho(0)\}
+
\mathrm{Tr}\{\hat B(0)\hat A(t)\hat \rho(0)\}
\Big)
&=
\int dx\, dp\;
W[\hat A(t)]\, W[\hat B(0)]\, W(x,p,0) \nonumber \\
&=
\int dx\, dp\;
A_W(t)\, B_W(0)\, W(x,p,0).
\end{align}

\section{Validity of a hybrid Wigner function}
\label{append:valid}

In this section, we discuss the validity of the hybrid Wigner function by showing the positivity of the joint density matrix. The positivity of a conditional density matrix $\hat \rho (t;\Gamma_c)$ is discussed in Ref.~\cite{antonov2025modeling}; nevertheless, for completeness, we also present the discussion here. 

For a given trajectory $\Gamma_c$, the density matrix at time $t$ can be expressed in terms of a map
\begin{align}
    \hat \rho(t;\Gamma_c) = \Lambda_{\Gamma_c} [\hat \rho(0) ]
.\end{align} 
We assume that the quantum master equation, Eq.~\eqref{eq:hybridQME}, is the Markovian limiting case of a Nakajima--Zwanzig master equation, valid when the bath's correlation time is much shorter than the characteristic time of the system~\cite{breuer2002theory}. 
Therefore, the generator, Eq.~\eqref{eq:hybridQME}, is chosen to be in Lindblad form, and the map $\Lambda_{\Gamma_c}$ is a CPTP map and $\hat \rho(t;\Gamma_c)$ is normalized and positive for a valid initial density matrix $\hat \rho (0)$.

Then, the density matrix conditioned on $\bm q_c$ is written as 
\begin{align}
    \hat \rho (\bm q_c,t) = \int \mathcal{D}[\Gamma_c] P[\Gamma_c] \hat \rho (t;\Gamma_c)\delta (\bm q_c -\bm q_c(t))
.\end{align}
The positivity of $\hat \rho (\bm q_c,t)$ is followed from the inequality  
\begin{align}
    \langle \phi | \hat \rho (\bm q_c,t) | \phi\rangle =  \int \mathcal{D}[\Gamma_c] P[\Gamma_c] \langle  \phi |\hat \rho (t;\Gamma_c)| \phi\rangle\delta (\bm q_c -\bm q_c(t)) \geq 0
,\end{align} which is proved by using the non-negativity of $P[\Gamma_c]$ and $ \langle  \phi |\hat \rho (t;\Gamma_c)| \phi\rangle$. After taking the Wigner transformation of $\hat \rho (\bm q_c,t)$, one obtains the hybrid Wigner function in the main text
\begin{align}
  W (x,p,\bm q_c,t )= \mathcal N \, \Xi [\hat \rho(\bm q_c,t) ]   
,\end{align}
where $\mathcal N$ is a normalization constant that depends on the dimensionality of the quantum state. 

An alternative approach is to start from a fully quantum description of subsystems and then take a semiclassical approximation for the degrees of freedom that are to be treated classically. 
Hybrid approaches of this type, where the Wigner transformation of some variable of the density matrix was performed, have been used in the literature; see, for instance, \cite{Javanainen1980Laser,Schutz2013Cooling}.

\section{Calculation of the two-time correlation function using a quasi-classical Langevin equation} 
\label{append:solveCorr} 
The Cartesian product for $\mathbf q(t_1)$ and $\mathbf q(t_2)$ is given by 
\begin{align}
    \mathbf q(t_1) \otimes\mathbf  q(t_2) 
    =& [ e^{-A t_1} \mathbf{q}(0) + \int_0^{t_1} ds_1\, e^{-A(t_1 - s_1)} B  \boldsymbol{\eta}(s_1)  ] 
    \otimes 
    [ e^{-A t_2} \mathbf{q}(0) + \int_0^{t_2}ds_2\, e^{-A(t_2 - s_2)} B  \boldsymbol{\eta}(s_2)  ]^T \nonumber \\
    =&e^{-A t_1} \mathbf{q}(0)\otimes \mathbf{q}(0) e^{-A^T t_2} 
    + \int_0^{t_1}ds_1\int_0^{t_2} ds_2 \,e^{-A(t_1 - s_1)} B  \boldsymbol{\eta}(s_1)  \otimes \boldsymbol{\eta}(s_2) B^T e^{-A^T(t_2 - s_2)} 
.\end{align}
After averaging over trajectories,
\begin{align}
    \langle \mathbf q(t_1) \otimes \mathbf q(t_2) \rangle 
    =&e^{-A t_1} \langle \mathbf{q}(0)\otimes \mathbf{q}(0)\rangle e^{-A^T t_2} 
    + \int_0^{{\rm min}(t_1,t_2)}ds_1\int_0^{{\rm min}(t_1, t_2)} ds_2 e^{-A(t_1 - s_1)} B  \langle \boldsymbol{\eta}(s_1)  \otimes \boldsymbol{\eta}(s_2) \rangle B^T e^{-A^T(t_2 - s_2)}.
\end{align}
The stochastic differential equation \eqref{eq:OULangevin} has the linear drift term, which represents an Ornstein-Uhlenbeck process. In this case, the vector $\mathbf{q}$ can be written as 
\begin{align}
\mathbf{q}(t) = e^{-A t} \mathbf{q}(0) + \int_0^t ds \,e^{-A(t - s)} B  \boldsymbol{\eta}(s)  
,\end{align}
and the matrix for the 2nd moment is given by
\begin{align}
\langle \mathbf q (t)\otimes \mathbf q (t)\rangle =  e^{-A t} \langle \mathbf q (0) \otimes \mathbf q(0)\rangle  e^{-A^T t}  
+  2\int_0^t ds \, e^{-A(t - s)} B B^T e^{-A^T(t - s)} 
.\label{eq:2ndMoment}
\end{align}
For $t_1 =t$ and $t_2 =0$, 
the two-time correlation function becomes 
\begin{align}
        \langle \bm q(t) \otimes \bm q(0) \rangle 
    =&e^{-A t} \langle \mathbf{q}(0)\otimes \mathbf{q}(0)\rangle
.\end{align}

\section{Heisenberg picture}
\label{append:HeisenClosed}
In the Heisenberg picture, the equation of motion for an operator $\hat O$ is given by
\begin{align}
\frac{d\hat O}{dt}
&=
\frac{i}{\hbar}\,[\hat H,\hat O]
+
\sum_k
\left(
\hat L_k^\dagger \hat O \hat L_k
-
\frac{1}{2}\left\{\hat L_k^\dagger \hat L_k,\hat O\right\}
\right).
\label{eq:lindblad_heisenberg}
\end{align}
The commutator is 
\begin{align}
    [\hat L_k^\dagger \hat L_k, \hat O ] = \hat L_k^\dagger [\hat L_k,\hat O] + [\hat L_k^\dagger,\hat O]\hat L_k.
\end{align}
The dissipator term can be written as 
\begin{align}
    \left(
\hat L_k^\dagger \hat O \hat L_k
-
\frac{1}{2}\left\{\hat L_k^\dagger \hat L_k,\hat O\right\}
\right) = \hat L_k^\dagger [\hat O, \hat L_k] - \frac{1}{2} \hat L^\dagger [\hat O,  \hat L_k] -\frac{1}{2}[\hat O,\hat L_k]\hat L^\dagger_k.
\end{align}
If the commutator $[\hat O,\hat L_k]$ is a $c$-number, the dissipative term does not generate higher-order operator products. More generally, if the action of the Lindbladian on a given operator set $\{\hat O_\alpha\}$ is closed within its linear span, the Heisenberg equations of motion constitute a closed set of coupled equations. 
\begin{align}
\hat O(t)
=
\sum_\alpha c_\alpha(t)\,\hat O_\alpha
+
c_0(t)\,\mathbb{I},
\end{align}
where the coefficients $c_\alpha(t)$ are time-dependent scalar functions. Consequently, the Heisenberg dynamics reduces to a finite-dimensional linear problem for these coefficients, and no additional operators are generated during the time evolution.

\section{Difference between two moments}
\begin{figure}
\includegraphics[width=0.48\textwidth]{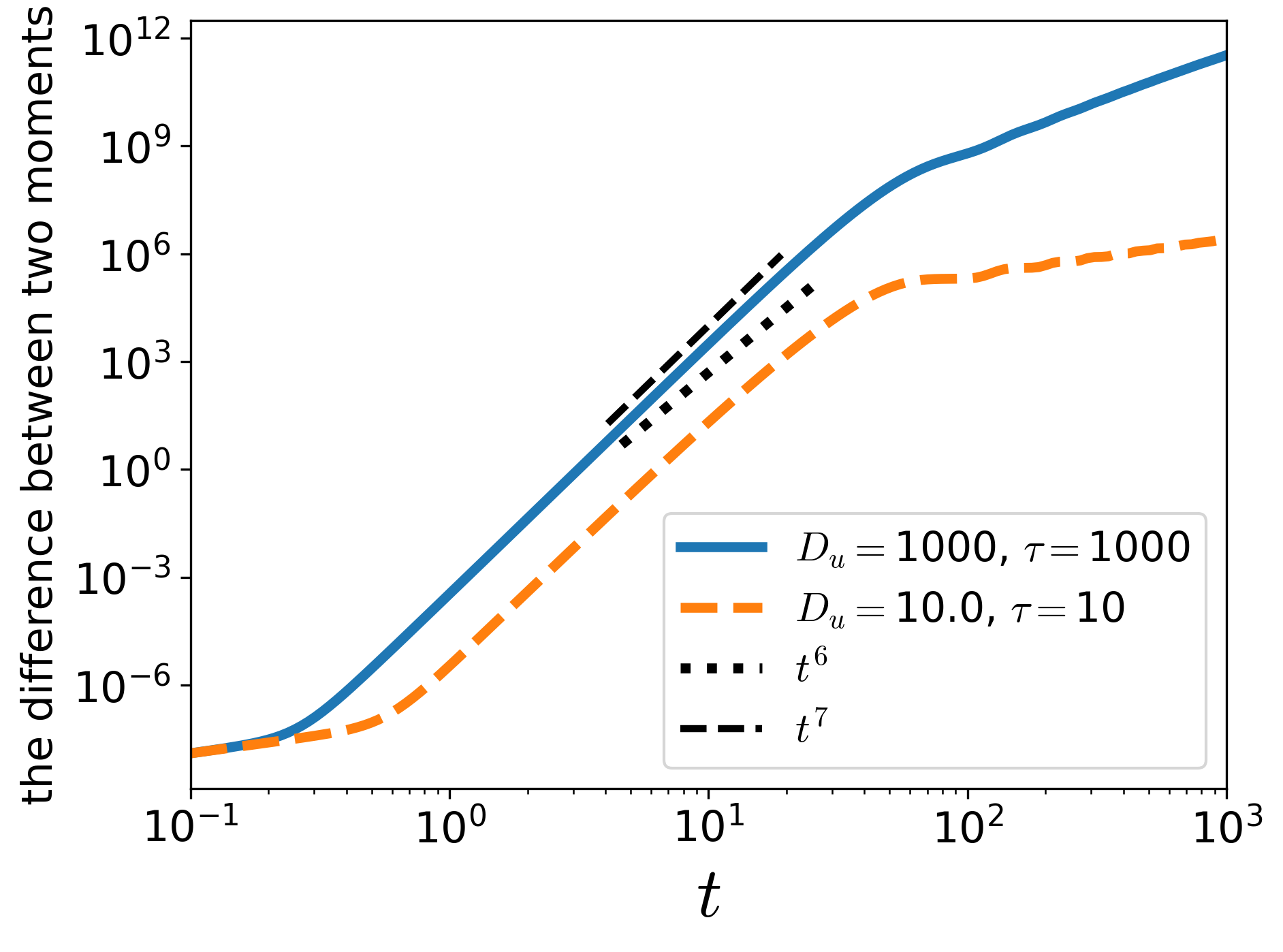}
\vskip -0.1in
\caption{ Plot of the difference between two moments, $\langle \hat x^2(t)\rangle_{ x_c} - \langle \hat x^2(0) \rangle_{x_c} $, versus time of the quantum active particle. 
The blue solid line represents the case of long persistence time ($\tau =1000$) and high intensity ($D_u=1000$). The orange solid line represents the case of shorter persistence time ($\tau=10$) and lower intensity ($D_u = 10$). 
The frequency of the quantum harmonic trap is chosen as $\omega=0.08$. The dissipation and pumping rates are $\nu_- = 10^{-4}$ and $\nu_+ = 10^{-8}$, respectively. We set the mass and Planck constant to unity.
}
\label{fig:twomomen}
\vskip -0.1in
\end{figure}

In the main text, we employ one definition of the MSD, whereas Ref. [21] uses another definition for the quantum MSD. 
Here, we examine whether the $t^{7}$ scaling persists under this alternative definition. The definition adopted in Ref.~\cite{antonov2025engineering} is given as follows
\begin{align}
    \text{(alternative MSD)} = \langle \hat x (t)^2 \rangle  - \langle \hat x (0)^2 \rangle 
\label{eq:alter_def},\end{align} which approximates $\langle \hat x(t) \hat x(0)\rangle + \langle \hat x(0) \hat x(t)\rangle \sim \langle \hat x^2(0)\rangle$

In Fig.~\ref{fig:twomomen}, we plot the quantum MSD based on the alternative definition using the same set of parameters as in the main text. We find that the $t^{7}$ slope remains under identical parameter conditions, showing that the $t^{7}$ scaling is robust with respect to the choice of MSD definition.

\section{Evaluating the two-time function}
\label{append:evaluateTwoTimeFunction}
\begin{figure}
\includegraphics[width=0.48\textwidth]{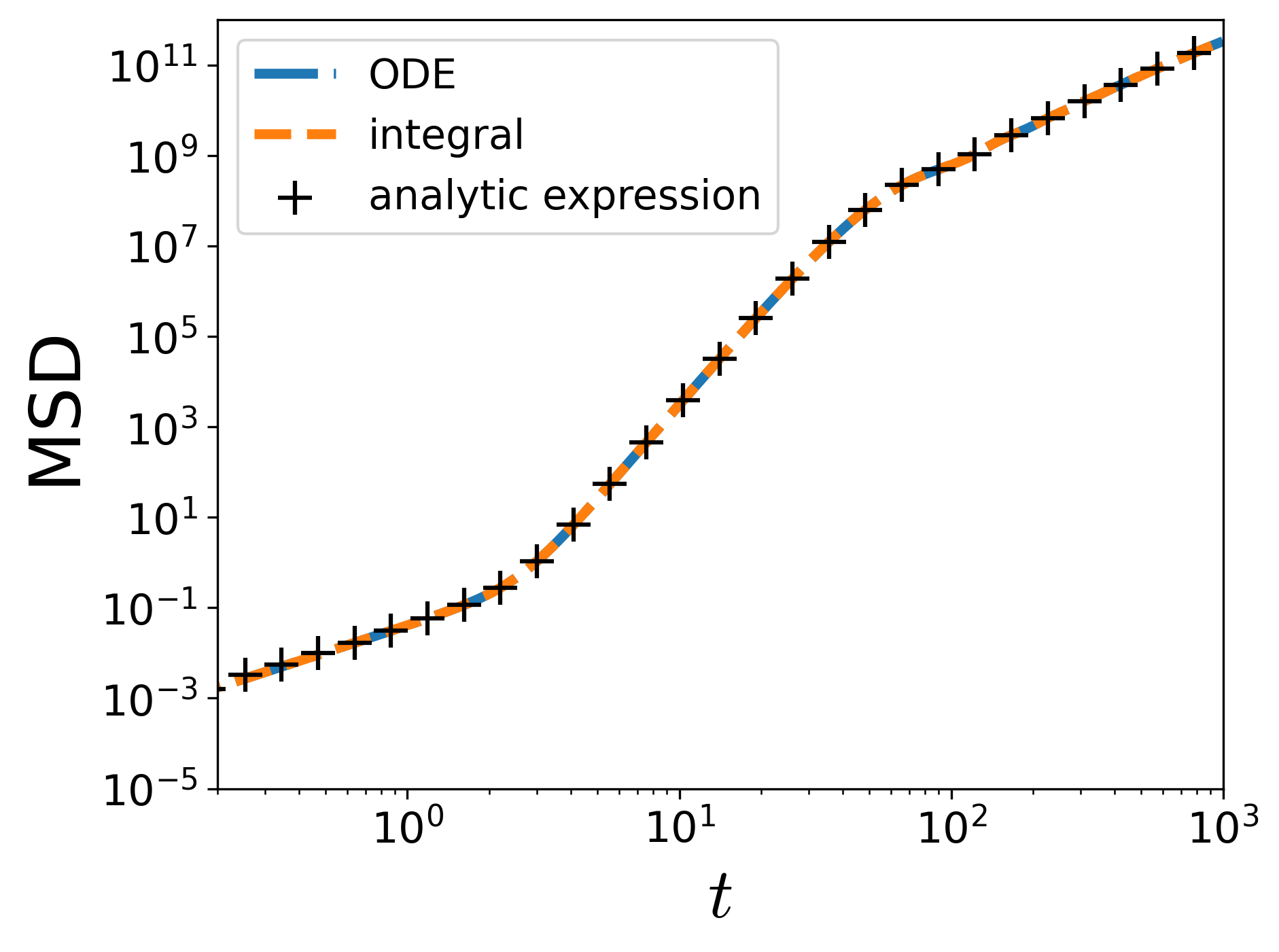}
\vskip -0.1in
\caption{Plot of MSD versus time of the quantum active particle calculated with three different methods. 
The blue line represents the result obtained by solving Eq.~\eqref{eq:ode}. The orange line represents the result obtaind by numerically evaluating integral expression, Eq.~\eqref{eq:Cov}. The cross symbol represents the analytically evaluated MSD; the explicit expression of MSD is omitted due to its complexity. 
The parameters all the same in Fig.~\ref{fig:t7t6activity} with $D_u=1000$ and $\tau=1000$. 
}
\label{fig:match}
\vskip -0.1in
\end{figure}

An alternative way to evaluate the two-time function in Eq.~\eqref{eq:Cov} is to solve the differential equation 
\begin{align}
   \partial_t \langle \mathbf q(t) \otimes \mathbf q(t)  \rangle =& -A \langle \mathbf q(t) \otimes \mathbf q(t)  \rangle -\langle \mathbf q(t) \otimes \mathbf q(t)  \rangle A^T+ 2B B^T
\label{eq:ode}\end{align} instead of directly evaluating the integrals in Eq.~\eqref{eq:Cov}.
We solved the above differential equation for plots in this paper. 
Another method is to numerically evaluate the analytical expression. In Fig.~\ref{fig:match}, we show that three methods agree within the given parameters. In the case of analytical expression, its precision is low due to many exponential expressions, particularly when $t \gg 10^3$.

\section{Hybrid Wigner master equation}
\label{append:hybridWigner}
For a given classical trajectory $\Gamma_c=\{x_c(t)\}$, the evolution of the conditional density matrix $\hat \rho_c$ 
is governed by
\begin{align}
\frac{d}{dt}\hat{\rho}_c(t)
=
-\frac{i}{\hbar}
\bigl[
\hat{H}(x_c(t)),\hat{\rho}_c(t)
\bigr]
+
\mathcal{D}_{x_c} \hat{\rho}_c(t).
\end{align}
Applying the Wigner transform, one may derive 
\begin{align}
\partial_t W(x,p,t | \Gamma_c )
=
\mathcal{L}_W 
W(x,p,t |  \Gamma_c),
\label{eq:fixed_Wigner}
\end{align}
where $\Gamma_c = \{ x_c(t), u_c(t) \}$ and $\mathcal L_W$ depends on current state $x_c(t)$ at time $t$. Then, we define a hybrid quasi-probability distribution which combines $x_c$ trajectory probability $P(\Gamma_c)$ and the conditional Wigner function $W(x,p,t | \Gamma_c )$
\begin{align}
W(x,p,x_c,u,t)
\equiv
\int \mathcal{D}[\Gamma_c]\,
W(x,p,t |  \Gamma_c) P[ \Gamma_c]
\delta \left(x_c - x_c(t)\right)
\delta \left(u - u_c(t)\right)
.\end{align} 
We call this function a hybrid quasi-probability distribution. After taking the time derivative of this object, one may derive 
\begin{align}
\partial_t W(x,p,x_c,u,t)
&=
\int \mathcal{D}[\Gamma_c]\,
\partial_t W(x,p,t | \Gamma_c)P[ \Gamma_c]
\delta(x_c-x_c(t))
\delta(u-u_c(t))
\nonumber \\
&\quad
+
\int \mathcal{D}[\Gamma_c]\,
W(x,p,t | \Gamma_c)P[ \Gamma_c]
\partial_t
\bigl[
\delta(x_c-x_c(t))\,
\delta(u-u_c(t))
\bigr]
\label{eq:mid_hybrid}.
\end{align}
Using Eq.~\eqref{eq:fixed_Wigner}, the first term on the right hand side of Eq.~\eqref{eq:mid_hybrid} can be expressed in terms of the operator $\mathcal L_W$. Since $\mathcal L_W$ depends on $x_c $ only at time $t$, the first term can be written as
\begin{align}
    \int \mathcal{D}[\Gamma_c]\,
\partial_t \,W(x,p,t | \Gamma_c)
\delta(x_c-x_c(t))
\delta(u-u_c(t)) = & \int \mathcal{D}[\Gamma_c]\,
\mathcal L_{W} W(x,p,t | \Gamma_c) P[ \Gamma_c]
\delta(x_c-x_c(t))
\delta(u-u_c(t)) \nonumber \\ 
= & \mathcal L_W W(x,p,x_c,u,t)
.\end{align}
The second term of Eq.~\eqref{eq:mid_hybrid} can be expressed with two terms as
\begin{align}
\int \mathcal{D}[\Gamma_c]\,
W(x,p,t | \Gamma_c) P[ \Gamma_c]
\partial_t
\bigl[
\delta(x_c-x_c(t))\bigr]
\delta(u-u_c(t))
\label{eq:x_cpart}\end{align}
and 
\begin{align}
\int \mathcal{D}[\Gamma_c]\,
W(x,p,t | \Gamma_c) P[ \Gamma_c]
\delta(x_c-x_c(t))
\partial_t
\bigl[
\delta(u-u_c(t))
\bigr]
\label{eq:u_cpart}.\end{align}
For each term, we follow a procedure similar to the Kramer-Moyal expansion~\cite{van1992stochastic}.
To this end, we introduce a test function~$\phi$ and consider the following expression: 
\begin{align}
\int d x_c \phi(x_c)\int &\mathcal{D}[x_c(t)]\,
W(x,p,t | \Gamma_c ) P[ \Gamma_c]
\partial_t [\delta(x_c-x_c(t)) ]\delta(u-u_c(t))\nonumber\\
&=
\sum_{n=1}^{\infty}
\frac{1}{n!}
\int dx_c\, \phi(x_c) 
\left(-\partial_{x_c}\right)^n
\int \mathcal{D}[\Gamma_c] \,
M_{n,x_c(t)}
W(x,p,t | \Gamma_c )P[ \Gamma_c]
\delta(x_c-x_c(t))\delta(u-u_c(t))\nonumber \\
&=
\sum_{n=1}^{\infty}
\frac{1}{n!}
\int dx_c \, [\partial_{x_c}^n \phi(x_c)]
M_{n,x_c }
\int \mathcal{D}[\Gamma_c ]
W(x,p,t | \Gamma_c ) P[ \Gamma_c]
\delta(x_c-x_c(t)) \delta(u-u_c(t)) \nonumber \\
&=
\sum_{n=1}^{\infty}
\frac{1}{n!}
\int d\Gamma_c\, \phi(x_c) 
\left(-\partial_{x_c}\right)^n
M_{n,x_c }
W(x,p,x_c,u,t)
\label{eq:midmxc}.\end{align}
Here, $M_{n,x_c(t)} = \lim_{\delta  t\rightarrow 0} {
\left[\delta  x_c(t)\right]^n}{\delta t}^{-1} $, which may be calculated using the Ito calculus and subsequently expressed in terms of $x_c(t)$. 
Because the $ M_{n,x_c }$ term in Eq.~\eqref{eq:midmxc} is a function of $x_c$, the term is independent of the integration variable $\Gamma_c$ of the path integral~$\int \mathcal D[\Gamma_c]$. 
After applying the same procedure to Eq.~\eqref{eq:u_cpart}, 
the summation of Eq.~\eqref{eq:x_cpart} and Eq.~\eqref{eq:u_cpart} is rearranged by $\mathcal L_{FP}W(x,p,x_c,u,t)$.
Finally, the following master equation is derived
\begin{align}
\partial_t W(x,p,x_c,u,t)
=
\mathcal L_W W(x,p,x_c,u,t)
+
\mathcal L_{FP} W(x,p,x_c,u,t)
.\end{align}

\section{Asymptotic form in the long time limit}
\label{appen:asymptotic}

\begin{figure}
\includegraphics[width=0.48\textwidth]{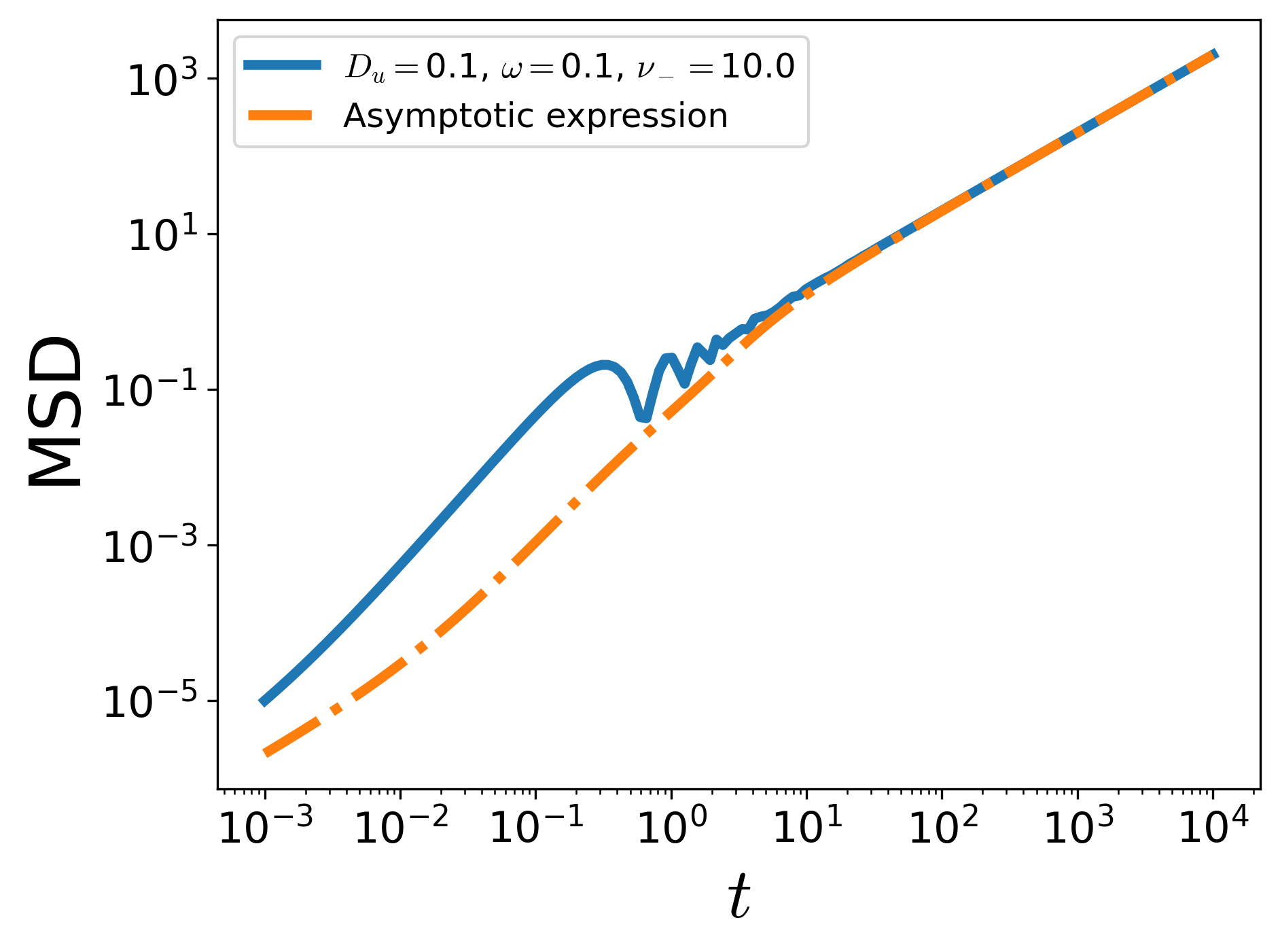}
\vskip -0.1in
\caption{ Plot of MSD versus time of the quantum active particle. $\tau=1$. The dash-dot line represents the asymptotic expression (Eq.~\eqref{eq:asymptotic_append}). The hierarchy between the three characteristic times is set to be $\tau \gg 1/\omega, 1/\gamma$. 
}
\label{fig:asymptotic}
\vskip -0.1in
\end{figure}

In the long time limit, the MSD of the quantum active matter system discussed in the main text [Eq.~\eqref{eq:hybrid_Lind}]  
can be written as
\begin{align}
\langle |\hat x(t) - \hat x(0)|^2 \rangle \sim 2D_u\tau^2 t + B_0
\Big[ 
-g_2(t) e^{-2t/\tau}
+g_1(t) e^{-t/\tau} 
- g_0
\Big]
\label{eq:asymptotic_append}.\end{align}
where 
\begin{align}
    B_0 =& \frac{D_u}{32(1+\tau^2\omega^2)^2},\\
    g_0 =& 16\tau^5\omega^2
\big(4 + \tau(4 + \tau(4+3\tau)\omega^2)\big),\\
g_1(t) =& 16\tau^5\omega^2(1+t+\tau)(4+4\tau^2\omega^2)
,\end{align}
and
\begin{align}
g_2(t) =& 16\tau^7\omega^4(2t+\tau)  
.\end{align}

The assumptions for the derivation are neglecting three terms that involves the initial condition in Eq.~\eqref{eq:Cov}. Since the initial condition information will be dissapeared in the long time limit. Also, we consider the hierarchy among characteristic times as $\tau \gg 1/\omega, 1/\gamma$. Under these assumptions, the MSD tensor can be rearranged by 
\begin{align}
    \langle (\mathbf q(t) -\mathbf q(0) ) \otimes (\mathbf q(t) -\mathbf q(0) ) \rangle \sim&
 2\int_0^t e^{-A(t - s)} B B^T e^{-A^T(t - s)} ds 
.\end{align}
$e^{-A (t-s)}$ term has four modes as the matrix $A$ has four eigenvalues $\lambda_i$ and their left and right eigenvectors $|R_i\rangle $ and  $|L_i\rangle $. Here $\langle L_i|R_j\rangle =\delta_{ij} $. The eigenvalues are $\lambda_1 =0$ $\lambda_2 = 1/\tau$, $\lambda_3 = \gamma/4 - i\omega  $ and $\lambda_4 = \gamma/4 + i\omega $. Because $\gamma$ is a large value, the eigen modes of $\lambda_3$ and $\lambda_4$ will decay rapidly in the long time limit. Thus,  
we neglecteigen modes of $\lambda_3$ and $\lambda_4$. 
 \begin{align}
 \langle (\mathbf q(t) -\mathbf q(0) ) \otimes (\mathbf q(t) -\mathbf q(0) ) \rangle  \sim&  2\int_0^t (|R_1\rangle \langle L_1| + |R_2\rangle \langle L_2| e^{-\lambda_2 (t-s)}  ) B B^T (|L_1\rangle \langle R_1| + |L_2\rangle \langle R_2| e^{-\lambda_2 (t-s)}  )
.\end{align}
After rearranging the (0,0) component of $ \langle (\mathbf q(t) -\mathbf q(0) ) \otimes (\mathbf q(t) -\mathbf q(0) ) \rangle $, the asymptotic expression, Eq.~\eqref{eq:asymptotic_append} can be derived. In Fig.~\ref{fig:asymptotic}, we plot Eq.~\eqref{eq:asymptotic_append}. In the long-time limit it shows good agreement. 

\section{Derivation of Eq.~\eqref{eq:t3t1}}
\label{append:deriv}
With the initial conditions $u (0) = 0$ and $x_c(0) = 0$, the MSD of $x_c$ (i.e., the MSD of an AOUP) in Eq.~\eqref{eq:t3t1} differs from the MSD in Eq.~\eqref{eq:usualAOUPMSD} that is obtained using the usual initial conditions, Eq.~\eqref{eq:inicond1}. 
To derive Eq.~\eqref{eq:t3t1}, we start from 
\begin{align}
    \dot x_c (t) = e^{-t/\tau } u (0) + \int^{t}_0 ds_2  e^{-\frac{1}{\tau } (t -s_2)}\sqrt{D} \eta_u(s_2),
\end{align}
where $ \langle  \eta(t)\eta(t')\rangle = 2 \delta (t-t') $.
Since $u(0)=0$, $x_c$ at time $t$ is written by 
\begin{align}
    x_c(t) = \int^t_0 ds_1 \int^{s_1}_0 ds_2  e^{-\frac{1}{\tau } (t -s_2)}\sqrt{D} \eta_u(s_2)
.\end{align}
Then, the MSD of $x_c$ is given by
\begin{align}
    \langle |x_c(t)- x_c(0)|^2  \rangle =& \int_0^t d s_1' \int_0^t d s_2' e^{-(s'_1+s'_2)/\tau }  \int ^{s'_1}_0  ds_1 \int^{s'_2}_0 ds_2 D e^{(s_1+s_2)/\tau }\langle \eta_u(s_1) \eta_u (s_2) \rangle \nonumber \\
    =& \int_0^t d s_1' \int_0^t d s_2' e^{-(s'_1+s'_2)/\tau }  \int ^{s'_1}_0  ds_1 \int^{s'_2}_0 ds_2 D e^{(s_1+s_2)/\tau }\delta(s_1 - s_2) \nonumber \\
    =& D_u \tau^2 [(-3 - e^{-2 t/\tau} + 4 e^{-t/\tau}) \tau + 2 t]
.\end{align}

\section{Adiabatic approximation criterion }
\label{append:criterion}

A standard adiabaticity criterion~\cite{albash2012quantum} for the validity of an adiabatic quantum master equation is given by
\begin{align}
    \ell \equiv \frac{h\hbar }{\Delta^2 t_f} \ll 1
\end{align} 
where
\begin{align}
    \Delta \equiv {\rm min}_{t\in [0,t_f],\, a,\, b} [\epsilon_1(t) -\epsilon_0(t)]\quad,\quad
    h\equiv {\rm max}_{t\in [0,t_f],\, a,\, b}|\langle \epsilon_a(t) |t_f\partial_t \hat H(t)  | \epsilon_b(t) \rangle |
,\end{align}
$t_f$ is the final time of a given process, $\epsilon_{0}$ and $\epsilon_1$ are the eigenvalues of the ground state and the first excited state of the Hamiltonian $\hat H$, respectively.

As the energy gap, $\epsilon_1 - \epsilon_0$, is time-independent, $\Delta$ is calculated by 
\begin{align}
    \Delta = \hbar \omega 
\label{eq:delta_cal}.\end{align}

By differentiating Eq.~\eqref{eq:hamiltonian} with respect to $t$ and using Eq.~\eqref{eq:CA1} and Eq.~\eqref{eq:u_def}, $h$ can be written as 
\begin{align}
    h =&  t_f\sqrt{\frac{\hbar m\omega^3}{2}}{\rm max}_{t\in [0,t_f],\, a,\, b}|\langle \epsilon_a(t) | (\hat a +\hat a^\dagger) u(t) | \epsilon_b(t) \rangle| \nonumber\\
    =&t_f\sqrt{\frac{\hbar m\omega^3}{2}}\sqrt{n_{\rm max} }{\rm max}_{t\in [0,t_f]} |u(t)| 
\label{eq:h_cal}.\end{align} Here, $n_{\rm max}$ is the maximum occupation number of our quantum system over the time interval $0\leq t\leq t_f$. 
Since the classical variable $x_c$ follows the dynamics in Eq.~\eqref{eq:AOUP}, while the quantum degrees of freedom are governed by Eq.~\eqref{eq:Lindblad-dynamic}, $\max_{t\in[0,t_f]}|u(t)|$ can be estimated from the steady-state distribution of $u$, while $n_{\rm max}$ can be estimated from the instantaneous stationary state:
\begin{align}
    {n_{\rm max}} \sim \max{\left(k_BT - \frac{\hbar \omega}{2},0\right)} \quad , \quad {\rm max}_{t\in [0,t_f]} |u(t)| \sim \sqrt{D_u\tau}
\label{eq:approx_nandu}.\end{align}
Using Eq.~\eqref{eq:approx_nandu}, Eq.~\eqref{eq:h_cal} can be rewritten as 
\begin{align}
    h \sim t_f\sqrt{\frac{ m\omega^2}{2}}\sqrt{ \max{\left(k_BT - \frac{\hbar \omega}{2},0\right)} } \sqrt{D_u\tau}
.\end{align}
Finally, this yields the criterion in the main text:
\begin{align}
    \ell \sim \sqrt{\frac{mD_u\tau}{2\omega^2\hbar^2} \max{\left(k_BT - \frac{\hbar \omega}{2},0\right)}} \ll 1
.\end{align}

\end{document}